\documentclass[12pt]{article}
\pdfoutput=1
\usepackage{jheppub}

\usepackage{verbatim}

\newcommand{\vac}{\left\vert \emptyset \right\rangle}
\newcommand{\CN}{{\cal N}}
\renewcommand{\c}{{\sf{c}}}
\renewcommand{\b}{{\sf{b}}}

\newcommand{\CM}{{\cal M}}

\newcommand{\CC}{{C}}
\newcommand{\CZ}{{Z}}
\newcommand{\CH}{{\cal H}}
\newcommand{\CP}{{\cal P}}

\newcommand{\C}{{\mathbb C}}
\newcommand{\bA}{{\mathbb A}}

\DeclareMathOperator{\Bun}{Bun}
\DeclareMathOperator{\Loc}{Loc}
\DeclareMathOperator{\Sb}{Sb}
\DeclareMathOperator{\Fc}{Fc}
\DeclareMathOperator{\Ff}{Ff}
\DeclareMathOperator{\Ai}{Ai}
\DeclareMathOperator{\Ext}{Ext}

\newcommand{\mc}{\mathcal}

\newcommand{\eps}{\epsilon}

\newcommand{\til}{\widetilde}

\newcommand{\Z}{\mathbb Z}

\newcommand{\op}{\operatorname}

\newcommand{\mbb}{\mathbb}

\newcommand{\R}{\mbb R}
\renewcommand{\d}{\mathrm{d}}

\title{Vertex Operator Algebras and 3d $\CN=4$ gauge theories}
\abstract{We introduce two mirror constructions of Vertex Operator Algebras 
associated to special boundary conditions in 3d $\CN=4$ gauge theories. 
We conjecture various relations between these boundary VOA's and 
properties of the (topologically twisted) bulk theories. We discuss applications to the
Symplectic Duality and Geometric Langlands programs.}

\author[1]{Kevin Costello,}
\author[1]{Davide Gaiotto}
\affiliation[1]{Perimeter Institute for Theoretical Physics, Waterloo, Ontario, Canada N2L 2Y5}

\begin{document}
\maketitle

\section{Introduction}
A recurring theme in supersymmetric gauge theory is the discovery of relations to the theory of 
Vertex Operator Algebras. Early examples can be found in four-dimensional, topologically twisted $\CN=4$ Super Yang Mills
\cite{Vafa:1994tf} and in $\Omega$-deformed four-dimensional $\CN=2$ gauge theory \cite{Nekrasov:2003rj,Alday:2009aq}.
All these examples can be understood by lifting the four-dimensional theories to
six dimensional SCFTs compactified on a Riemann surface, which provide the ``ambient space'' for the VOAs.

The idea that VOAs can be embedded into the algebra of local operators in a higher-dimensional quantum field theory 
can be generalized beyond the six-dimensional setting \cite{Beem:2014kka}. Indeed, certain protected correlation functions in 
superconformal field theories are encoded in VOA's \cite{Beem:2013aa,Beem:2014aa}. Furthermore, the six-dimensional setup can be 
mapped to configurations involving junctions of boundary conditions in topologically twisted $\CN=4$ Super Yang Mills \cite{Gaiotto:2017euk}.

In all of these examples, the VOAs live in the physical space of the quantum field theory. They encode algebras of 
local operators which are decoupled from the rest of the theory either by supersymmetry considerations or by an explicit 
topological twist of the theory. 

In this paper we present a construction of VOAs in three-dimensional $\CN=4$ gauge theories. The VOAs emerge as algebras of 
local operators at special, holomorphic boundary conditions for the topological twist of the bulk theory. They are very much analogous to the 
RCFTs which can be found at holomorphic boundary conditions of ordinary Chern-Simons theories.

The original motivation for introducing these VOAs is that they can provide a powerful computational tool 
to study the bulk TFT. For example, they may make manifest IR symmetries of the theory, which would be 
hard to account for with traditional methods \cite{Gaiotto:2016wcv} but are necessary for certain applications, 
such as the gauge theory interpretation of the Geometric Langlands program \cite{Kapustin:2006pk,Gaiotto:2008ak,Gaiotto:2016aa}.

In this paper we will find further mathematical motivations, including relations to the Symplectic Duality program. 
Information may also flow in the opposite direction, as gauge theory constructions provide a new framework to 
understand, organize and predict a variety of results in the theory of VOAs \cite{Creutzig:2017uxh}.  

\subsection{Structure of the paper}
In Section \ref{sec:bc} we discuss the definition of the holomorphic boundary conditions we employ.
In Section \ref{sec:bulkboundary} we discuss the relation between properties of the bulk TFT and of the 
boundary VOA. In Sections \ref{sec:H} and \ref{sec:C} we give a concrete definition of the two classes of boundary
VOAs, and verify in some simple examples that bulk theories have isomorphic boundary VOAs.  We conclude with some extra open problems.

\section{Deformable $(0,4)$ boundary conditions.}\label{sec:bc}
\subsection{Generalities}
Supersymmetric quantum field theories can be {\it twisted} by passing to the cohomology of a nilpotent 
supercharge $Q$, i.e. by adding the nilpotent supercharge to the BRST charge of the theory \cite{Witten:1988ze,Witten:1988xj}. 

The remaining supercharges $Q_a$ of the theory play an important role: the anti-commutator 
\begin{equation}
\{Q,Q_a\} = c_a^\mu P_\mu
\end{equation}
will make some of the translation generators $Q$-exact. 

Correlation functions of Q-closed operators will remain unchanged if any of the operators are translated in these directions. 
If the right hand side of the anti-commutator is a real translation generator, the theory will be topological in that direction. 
If the right hand side is a complex combination of two translation generators, the theory may be instead holomorphic 
in the corresponding plane. 

Similar considerations apply to BPS defects of the SQFT. Any defect which preserves $Q$ will survive as a defect in the twisted theory. 
As the defect will break some of the other supercharges $Q_a$, the topological or holomorphic properties of the 
defect local operators may differ from these of the bulk local operators. 

In particular, one may have holomorphic defects within a topological bulk theory. Our objective is to build holomorphic 
boundary conditions and interfaces for topological twists of three-dimensional $\CN=4$ quantum field theories.
Such defects will support Vertex Operator Algebras of holomorphic local operators.

We will quickly demonstrate that this objective cannot be accomplished by twisting 
any standard Lorentz-invariant BPS boundary conditions for $\CN=4$ theories. Instead, we will 
follow a more circuitous route.

\subsection{Nilpotent supercharges}
Consider at first the 3d supersymmetry algebra in the absence of central charges: 
\begin{equation}
\{Q^i_\alpha, Q^j_\beta\} = \delta^{ij} P_{\alpha \beta}
\end{equation}
where latin indices $i$,$j$ run from $1$ to $\CN$ and label the different sets of supercharges, while greek indices label the two
spinor components. 

The superalgebra admits nilpotent supercharges. They take the form $n^\alpha_i Q^i_\alpha$ 
with $\sum_i n^\alpha_i n^\beta_i=0$. This means that the two vectors $n^1_i$ and $n^2_i$ generate a null plane or a null line in $\C^\CN$. 

The corresponding exact translations take the form $n^\alpha_i P_{\alpha \beta}$. If $n^1_i$ and $n^2_i$ are collinear, 
the twist is holomorphic. Without loss of generality we can take the exact generators to be translations in the $x^3$ direction 
and anti-holomorphic derivatives in the $x^{1,2}$ plane. Otherwise, the twist is topological. 

If $\CN=1$, no twists are possible. If $\CN=2$, the only possible twists are holomorphic. If $\CN=4$, 
the 2-form $\epsilon_{\alpha \beta} n^\alpha_i n^\beta_j$ may either be self-dual, anti-self-dual or vanish. 
We denote the corresponding families of nilpotent supercharges as H-type, C-type or holomorphic supercharges. 

We are mostly interested in the $\CN=4$ case. Up to discrete identifications, the R-symmetry group is conventionally denoted as 
$SU(2)_H \times SU(2)_C$. The eight supercharges can be correspondingly denoted as $Q^{A \dot A}_\alpha$, 
with all types of indices running over $1,2$ and SUSY algebra 
\begin{equation}
\{Q^{A \dot A}_\alpha, Q^{B \dot B}_\beta\} = \epsilon^{AB} \epsilon^{\dot A \dot B} P_{\alpha \beta}
\end{equation}

Up to complexified Lorentz transformations, generic H-type, C-type and holomorphic supercharges take the form 
\begin{equation}
\tilde \zeta_{\dot A} \delta^\alpha_A Q^{A \dot A}_\alpha \qquad \qquad \zeta_{A} \delta^\alpha_{\dot A} Q^{A \dot A}_\alpha \qquad \qquad \zeta_{A} \tilde \zeta_{\dot A} Q^{A \dot A}_+
\end{equation}

In particular, theories with unbroken $SU(2)_C$ admit a fully topological twist where the Lorentz group is twisted by $SU(2)_C$ 
to produce a scalar, C-type supercharge. This is the analogue of the Rozansky-Witten twist for $\CN=4$ sigma models \cite{Rozansky:1996bq}. 
The parameter $\zeta_{A}$ is a choice of complex structure on the Higgs branch of the theory. 

Similarly, theories with unbroken $SU(2)_H$ admit a fully topological twist where the Lorentz group is twisted by $SU(2)_H$ 
to produce a scalar, H-type supercharge. This is the mirror of the Rozansky-Witten-like twist \cite{Kapustin:2010ag}.
The parameter $\tilde \zeta_{\dot A}$ is a choice of complex structure on the Coulomb branch of the theory. 

It is useful to think about the C- and H-twists as small deformations of the holomorphic twist. 
For example, the holomorphic supercharge $Q^{+ \dot +}_+$ can be deformed to H- and C- type supercharges 
\begin{equation}
Q^{+ \dot +}_+ + \epsilon Q^{- \dot +}_- \qquad \qquad Q^{+ \dot +}_+ + \epsilon Q^{+ \dot -}_- 
\end{equation}

As the $\epsilon$ parameters are charged under Lorentz transformations, perturbation theory 
in $\epsilon$ will often be exact. 

\subsection{Deformations of boundary conditions}

Physical, Lorentz-invariant BPS boundary conditions (or interfaces) for a 3d SQFT will preserve some 
collections of $\CN_\pm$ supercharges with positive and negative chirality in the plane parallel to the boundary.
The basic constraint is that the preserved supercharges should not anti-commute to translations
in the direction perpendicular to the boundary. Hence the supercharges of positive and negative 
chiralities should span two orthogonal subspaces $V_\pm$ of $\R^\CN$. 

The preserved supercharges will form an $(\CN_-, \CN_+)$ 2d superalgebra. We denote the corresponding 
class of boundary conditions as $(\CN_-, \CN_+)$ boundary conditions. 

Such boundary conditions will be compatible with a topological twist only if $n^i_\pm$ belongs to $V_\pm$. 
As these vectors are null, we need the boundary condition to preserve at least two supercharges of each chirality. 
For $\CN=4$, that means $(2,2)$ boundary conditions. These are interesting, well studied boundary conditions \cite{Kapustin:2008sc,Kapustin:2009uw,Bullimore:2016nji}, 
but it is easy to see that the bulk topological twist makes $(2,2)$ boundary conditions topological as well.

In order to find interesting boundary VOAs, we clearly need to look at non-Lorentz invariant boundary conditions. 
On the other hand, in order to make contact with dualities and other non-perturbative results we should 
not stray far from physical, Lorentz-invariant BPS boundary conditions. 

Based on previous work on a variety of examples \cite{Gaiotto:2016wcv,Gaiotto:2017euk}, our compromise will be to look for some canonical deformations 
of physical boundary conditions preserving $(0,4)$ supersymmetry, which are another interesting 
class of half-BPS boundary conditions which have interesting duality properties \cite{Chung:2016pgt, Dimofte:2017tpi}.
These boundary conditions are obviously compatible with a bulk holomorphic twist, 
as they preserve all supercharges with positive chirality. 

Concretely, the statement that the boundary condition breaks the anti-chiral supercharges means that 
the restriction to the boundary of the normal component of the corresponding supercurrents $S^{A \dot A}_{-,\mu}$ is not a total derivative. 

Consider a small deformation of a generic $(0,4)$ boundary condition by some boundary operator $O$:
\begin{equation}
\epsilon \int_\partial d^2x O(x)
\end{equation}
Such a deformation will break the holomorphic supersymmetries $Q^{A \dot A}_+$ if 
$Q^{A \dot A}_+ O$ is not a total derivative. Concretely, that means that the restriction to the boundary of the 
normal component $S^{A \dot A}_{+,\perp}$ of the corresponding supercurrent  does not vanish
after the deformation, but equals $Q^{A \dot A}_+ O$.

On the other hand, the deformed boundary condition will preserve the deformed H-type supercharge 
$Q^{+ \dot +}_+ - \epsilon Q^{- \dot +}_-$ if we can arrange for 
\begin{equation}\label{eq:obH}
S^{- \dot +}_{-,\perp}|_\partial = Q^{+ \dot +}_+ O
\end{equation}

As long as $Q^{- \dot -}_+$ remains (or can be deformed to) a symmetry of the deformed boundary condition, 
then the twisted, deformed boundary condition will be holomorphic. 

Similarly, if $S^{+ \dot -}_{-,\perp} = Q^{+ \dot +}_+ O$ we can deform the boundary condition 
to make it compatible with a bulk topological twist based on $SU(2)_C$.

For the examples we will study in this paper, one can laboriously check in the physical theory
by hand that the desired deformation exists. A simpler strategy is to first pass to the holomorphic twist of the 
theory and boundary conditions and then work out the obstruction within the twisted theory.
We do so in a companion paper \cite{Companion}

\subsection{Example: free hypermultiplet}
There are two natural $(0,4)$ boundary conditions for a free hypermultiplet: Neumann and Dirichlet.
The terminology is associated to the boundary conditions for the four real scalars in the hypermultiplet. 
The fermion boundary conditions are then determined by supersymmetry. 

The boundary conditions and their deformations are discussed briefly in Appendix E of \cite{Gaiotto:2017euk}.
The result is that:
\begin{itemize}
\item Neumann b.c. can be deformed to be compatible with an $SU(2)_H$ twist. 
The resulting boundary condition supports the VOA of {\it symplectic bosons}, which we denote as $\Sb$. 
\item Dirichlet b.c. can be deformed to be compatible with an $SU(2)_C$ twist. 
The resulting boundary condition supports the VOA of {\it fermionic currents},
basically a $\mathrm{psu}(1|1)$ Kac-Moody VOA, which we denote as $\Fc$. 
\end{itemize}
There are two intuitive ways to understand these results. 

The $SU(2)_C$, or Rozansky-Witten, twist
of free hypers is known to give a fermionic version of Chern-Simons theory, with the 
symplectic form playing the role of Chern-Simons coupling. Dirichlet boundary conditions in such 
a Chern-Simons theory naturally produce a fermionic current algebra \cite{Witten:1988hf}.

On the other hand, the $SU(2)_H$ twist of free hypers gives a precisely the bulk theory which 
controls the analytic continuation of a two-dimensional path-integral, in the sense of 
\cite{Witten:2010cx, 2016arXiv161100311B}, for the symplectic boson action \cite{Gaiotto:2016wcv}
\begin{equation}
\int d^2x \langle Z, \bar \partial Z\rangle
\end{equation}
The deformed Neumann boundary condition are precisely the boundary conditions whose local operator algebra coincides with observables 
for the symplectic boson path integral. 

\subsection{Example: free vectormultiplet}

We expect $(0,4)$ Neumann and Dirichlet boundary conditions for a general pure $U(1)$ gauge theory 
to admit deformations compatible respectively with an H- and a C-twists. This should follow, for example, 
from the mirror symmetry relation between free $U(1)$ gauge fields and a free hypermultiplet 
valued in $S^1 \times \R^3$. 

Dirichlet boundary conditions will support boundary monopole operators, whose quantum numbers and properties 
depend on the bulk and boundary matter fields. These operators will give important contributions to the boundary VOAs
but are non-perturbative in nature and require a careful analysis.

Neumann boundary conditions, instead, do not support boundary monopole operators and 
the corresponding VOAs are simpler to understand. 

The supersymmetry transformation of an Abelian vectormultiplet are schematically 
\begin{align}
Q^{A \dot A}_\alpha A_{\beta \gamma} &= \epsilon_{\alpha (\beta)} \lambda^{A \dot A}_{\gamma)} \cr
Q^{A \dot A}_\alpha \Phi^{\dot B \dot C} &= \epsilon^{\dot A (\dot B)} \lambda^{A \dot C}_\alpha \cr
Q^{A \dot A}_\alpha \lambda^{B \dot B}_{\beta} &= F_{\alpha \beta} \epsilon^{A  B}\epsilon^{\dot A \dot B} + \epsilon^{A  B} \partial_{\alpha \beta} \Phi^{\dot A \dot B}
\end{align}
The supercurrents are  schematically 
\begin{equation}
S^{A \dot A}_{\alpha \beta \gamma} = F_{(\alpha \beta} \lambda^{A \dot A}_{\gamma)} + \epsilon_{\dot B \dot C}  \partial_{(\alpha \beta} \Phi^{\dot A \dot B}\lambda^{A \dot C}_{\gamma)}
\end{equation}
A $(0,4)$ boundary condition must satisfy at the boundary $S^{A \dot A}_{++-}=0$.

Neumann boundary conditions for the gauge fields require 
\begin{equation}
F_{++} = F_{--} =0 \qquad \qquad \lambda^{A \dot A}_{+}=0 \qquad \qquad \partial_{++}\Phi^{\dot A \dot B} = \partial_{--}\Phi^{\dot A \dot B}=0
\end{equation}
and in particular impose Dirichlet b.c. for the vectormultiplet scalars. 

The normal component $S^{- \dot +}_{-,\perp}= S^{- \dot +}_{+--}$ becomes 
\begin{equation}
F_{+-}\lambda^{- \dot +}_{-} + \epsilon_{\dot B \dot C} \partial_{+-} \Phi^{\dot + \dot B}\lambda^{- \dot C}_{-}= Q^{+ \dot +}_+ \left( \epsilon_{\dot B \dot C} \lambda^{- \dot B}_- \lambda^{- \dot C}_{-}  \right)
\end{equation}
suggesting that a deformation compatible with H-twist is possible, as expected. \footnote{On the other hand, the normal component $S^{+ \dot -}_{-,\perp}= S^{+ \dot -}_{+--}$ becomes 
\begin{equation}
F_{+-}\lambda^{+ \dot -}_{-} + \epsilon_{\dot B \dot C} \partial_{+-} \Phi^{\dot - \dot B}\lambda^{+ \dot C}_{-}
\end{equation}
which does not seem to be $Q^{+ \dot +}_+$-exact.}

If we keep the same bosonic boundary conditions and deform the fermion boundary condition $\lambda^{+ \dot A}_{+}=0$ to $\lambda^{+ \dot A}_{+} + \lambda^{- \dot A}_{-}=0$, then at the boundary $S^{+ \dot A}_{++-} + S^{- \dot A}_{+--}$ vanishes and we have an H-twist compatible Neumann boundary condition. 

Dirichlet b.c. for the gauge fields require 
\begin{equation}
F_{+-} =0 \qquad \qquad \lambda^{A \dot A}_{-}=0 \qquad \qquad \partial_{+-}\Phi^{\dot A \dot B} = 0
\end{equation}
and in particular impose Neumann b.c. for the vectormultiplet scalars. 

On the other hand, the normal component $S^{+ \dot -}_{-,\perp}= S^{+ \dot -}_{+--}$ becomes 
\begin{equation}
F_{--}\lambda^{+ \dot -}_{+} + \epsilon_{\dot B \dot C} \partial_{--} \Phi^{\dot - \dot B}\lambda^{+ \dot C}_{+} = Q^{+ \dot +}_+ \left( F_{--} \Phi^{\dot - \dot -} + \Phi^{\dot - \dot -} \partial_{--} \Phi^{\dot - \dot +}\right) + \partial_{--} \cdots
\end{equation}

That indicates the existence of a deformation compatible with C-twist, which changes the boundary conditions for the bosons and leaves the boundary conditions for the fermions unchanged, as expected. 

\subsection{Index calculations}
Supersymmetric localization allows for the calculation of non-trivial Witten indices of spaces of local operators 
in 3d SQFTs with at least $\CN=2$ SUSY \cite{Imamura:2011su,Kapustin:2011jm,Gadde:2013wq,Gadde:2013sca,Dimofte:2017tpi}. 
These indices essentially compute the Euler character of the spaces of local operators compatible with an holomorphic twist, 
weighed by fugacities for the symmetries which commute with the holomorphic super-charge. 
There is a supersymmetric index which counts protected 
bulk local operators and a half-index which counts local operators at $(0,2)$ boundary conditions. 

There are two important specializations of the index or half-index of $\CN=4$ systems, which restrict the fugacities 
to symmetries preserved by either H- or C- topological super-charges and thus compute the Euler character of the spaces of
local operators compatible with the corresponding twist. This is true even for deformed $(0,4)$ boundary conditions, as the index 
is insensitive to the deformation. 
 
In practice, that means half-index calculations give us access to the characters of the vacuum module of 
the boundary VOAs we seek.

\subsubsection{Example: hypermultiplet indices}
The bulk index of a single chiral multiplet of fugacity $x$, in appropriate conventions, is 
\begin{equation}
I_{ch}(x;q) = \frac{(q x^{-1};q)_\infty}{(x;q)_\infty} = \prod_{n \geq 0}\frac{1-x^{-1} q^{n+1} }{1-x q^n}
\end{equation}
This index simply counts words made out of derivatives of the chiral multiplet complex scalar and 
one of the fermions in the multiplet. The $q$ fugacity measures a combination of spin and R-charge. 

Half-indices for Neumann or Dirichlet boundary conditions include only one tower: 
\begin{align}
II_{ch, N}(x;q) &= \frac{1}{(x;q)_\infty} = \prod_{n \geq 0}\frac{1}{1-x q^n} \cr
II_{ch,D}(x;q) &= (q x^{-1};q)_\infty = \prod_{n \geq 0}(1-x^{-1} q^{n+1})
\end{align}

The index for a full hypermultiplet combines to chiral multiplets:
\begin{equation}
I_{hyper}(x;y;q) = I_{ch}(x y;q)I_{ch}(x^{-1} y;q) = \frac{(q x y^{-1};q)_\infty(q x^{-1} y^{-1};q)_\infty}{(x y;q)_\infty(x^{-1} y;q)_\infty} 
\end{equation}
The H-twist restricts the fugacities by $y = q^{\frac12}$. The resulting index is precisely $1$: the free hypermultiplet 
has no ``Coulomb branch local operators'', which would survive in the H-twist. 

On the other hand, the C-twist restricts the fugacities by $y = 1$. The index becomes simply $(1-x)^{-1}(1-x^{-1})^{-1}$,
with the two factors corresponding to the two generators of the algebra of Higgs branch local operators, $\C[u,v]$. 

The half-index for a typical $(2,2)$ boundary condition, setting to zero at the boundary one of the two complex scalars in the hypermultiplet, 
takes the form 
\begin{equation}
II_{hyper, (ND)}(x;y;q) = II_{ch, N}(x;q)II_{ch,D}(x^{-1};q)) = \frac{(q x y^{-1};q)_\infty}{(x y;q)_\infty} 
\end{equation}
As we specialize $y = 1$ or $y=q^{\frac12}$, the half-index again simplify drastically, as expected for a topological boundary condition.

The half-index for a Neumann $(0,4)$ boundary condition takes the form 
\begin{equation}
II_{hyper, (NN)}(x;y;q) = II_{ch, N}(x;q)II_{ch,N}(x^{-1};q)) = \frac{1}{(x y;q)_\infty(x^{-1} y;q)_\infty} 
\end{equation}
If we restrict fugacities according to the H-twist we obtain the vacuum character for the symplectic boson VOA
\begin{equation}
II_{hyper, (NN)}(x;q^{\frac12};q) = \frac{1}{(x q^{\frac12};q)_\infty(x^{-1} q^{\frac12};q)_\infty} =\chi_{\Sb}(x;q)
\end{equation}

The half-index for a Dirichlet $(0,4)$ boundary condition takes the form 
\begin{equation}
II_{hyper, (DD)}(x;y;q) = II_{ch, D}(x;q)II_{ch,D}(x^{-1};q)) = (q x y^{-1};q)_\infty(q x^{-1} y^{-1};q)_\infty 
\end{equation}
If we restrict fugacities according to the C-twist we obtain the vacuum character for the fermionic current VOA
\begin{equation}
II_{hyper, (DD)}(x;1;q) =(q x ;q)_\infty(q x^{-1} ;q)_\infty  =\chi_{\Fc}(x;q)
\end{equation}

\subsubsection{Example: vectormultiplet half-indices}
The half-index for a $U(1)$ $\CN=2$ gauge multiplet with Neumann $(0,4)$ boundary conditions and no charged matter is simply 
\begin{equation}
II_{gauge, N}(q) = (q;q)_\infty
\end{equation}

The half-index for a $U(1)$ $\CN=4$ gauge multiplet with Neumann $(2,2)$ boundary conditions and no charged matter is  
\begin{equation}
II_{vector, NN}(q) = II_{gauge, N}(q) II_{chiral, N}(q y^{-2};q)= \frac{(q;q)_\infty}{(q y^{-2};q)_\infty}
\end{equation}
As expected, most factors cancel out in the denominator both for $y = 1$ or $y=q^{\frac12}$: 
for the C-twist everything cancels out and is trivial and for the H-twist one is left with a divergent factor counting 
topological local operators made out of polynomials in a single field with no fugacity. 

The half-index for a $U(1)$ $\CN=4$ gauge multiplet with Neumann $(0,4)$ boundary conditions and no charged matter is  
\begin{equation}
II_{vector, ND}(q) = II_{gauge, N}(q) II_{chiral, D}(q y^{-2};q)= (q;q)_\infty (y^{2};q)_\infty
\end{equation}
In the H-twist we get $(q;q)_\infty^2$, from the two fermionic local operators which survive at the boundary.
Later on, we will identify them with operators annihilated by $b_0$ in a $b c$ VOA of ghosts 
for a 2d chiral gauge theory.

In the absence of matter fields, the boundary monopole operators at Dirichlet boundary conditions have no 
spin or R-symmetry charge. They only carry integral charges for the bulk ``topological'' $U(1)$ gauge symmetry. 
In each topological charge sector, 
\begin{equation}
II_{gauge, D,n}(q) = (q;q)^{-1}_\infty
\end{equation}

The analysis is similar as before. For the $(0,4)$ Dirichlet boundary conditions we get 
\begin{equation}
II_{vector, DN,n}(q) = (q;q)^{-1}_\infty(q y^{-2};q)^{-1}_\infty
\end{equation}
In the C-twist we get $(q;q)^{-2}_\infty$ in each charge sector. 

Somewhat formally, this  is compatible with the 
expectation from mirror symmetry to the H-twist of a hypermultiplet 
valued in $\C \times \C^*$: a $\beta \gamma$ system with $\gamma$ valued in $\C^*$. 
\footnote{Indeed, if we ignore the $\gamma$ zeromodes, the $\beta \gamma$ vacuum character 
would be $(q;q)^{-1}_\infty(q t;q)^{-1}_\infty$, where the $t$ fugacity counts the $U(1)$ 
charge carried by $\gamma$. Expanding that out into powers of $t$, and 
adding together the contributions from operators of charge $k$ multiplying $\gamma^{n-k}$ 
gives back $(q;q)^{-1}_\infty(q;q)^{-1}_\infty$. }

\section{Boundary conditions and bulk observables} \label{sec:bulkboundary}

\subsection{Boundary VOA and conformal blocks}
Conformal blocks for a VOA are essentially defined as collections of ``correlation functions''
of VOA local operators on some Riemann surface $C$ which are consistent with OPE. 
\footnote{In the current setup, and in general in any situation involving non-unitary, cohomological field theories, 
conformal blocks should be intended in a derived sense: a proper calculation may result in 
unexpected contributions in non-trivial ghost numbers, which can play important roles when conformal blocks are 
manipulated. For example, if we build conformal blocks through a sewing construction, 
gluing trinions together into a Riemann surface, the gluing procedure involves tensor products 
over the VOA. These tensor products should be intended as derived tensor products. }

If we take the twisted 3d gauge theory on a geometry of the form $\R^+ \times C$, 
inserting local operators at the boundary and some asymptotic state at infinity for the TFT, 
we get precisely such a consistent collection. That means we always have a map from the 
Hilbert space of the 3d TFT compactified on $C$ to the space of conformal blocks for 
any boundary VOA. 

Such a map is often an isomorphism. This statement becomes more likely to be true 
if we account for the fact that the map is not just a map of vector spaces (or better, complexes):
as we vary the complex structure of $C$, both conformal blocks and the Hilbert space 
describe matching flat bundles (or better, D-modules) over the moduli space of complex structures. 

The relation between the TFT Hilbert space and the VOA conformal blocks was an important motivation for this work: 
physical constructions relevant for Geometric Langland involve 3d TFTs which do not admit a complete Lagrangian description, 
but have known boundary VOAs. The study of conformal blocks of these VOAs gives access to otherwise unavailable 
information about the TFT Hilbert spaces. 

\subsection{Bulk operator algebra and $\Ext$ groups}
The space of bulk local operators for the TFT is closely related to the Hilbert space of states on a two-sphere. 
That means the VOA should also give access to the space of bulk local operators.

We conjecture that the algebra of bulk local operators can be described as the self-$\Ext$ groups of the vacuum module of the boundary VOA.   This is one of the main conjectures in this paper. As we will see, it is a rather non-trivial statement. 
For example, it will allow us to recover the recent mathematical definition \cite{Braverman:2016wma} of the algebra of Coulomb branch local operators
of 3d $\CN=4$ gauge theories. 

Let us explain heuristically why we expect this to be true.  Let $\mc{O}_{\text{bulk}}$ denote the space of bulk local operators, and $\mc{O}_{\text{boundary}}$ the space of boundary local operators.   The space of bulk operators acts on the space of boundary operators, using the OPE between bulk and boundary operators. This action gives an algebra homomorphism map
\begin{equation} 
\mc{O}_{\text{bulk}} \to \op{End}(\mc{O}_{\text{boundary}}). 
 \end{equation}
 The algebra of boundary charges\footnote{The algebra of charges is a topological associative algebra with the property that modules for this associative algebra are the same as modules for the vertex algebra we start with. }    -- generated by countour integrals of boundary local operators -- also acts on the space of boundary local operators. If we denote the algebra of boundary charges by $\oint \mc{O}_{\text{boundary}}$, we have a homomorphism
 \begin{equation} 
 \oint \mc{O}_{\text{boundary}} \to \op{End}(\mc{O}_{\text{boundary}}). 
  \end{equation}

The actions of $\mc{O}_{\text{bulk}}$ and $\oint \mc{O}_{\text{boundary}}$ on $\mc{O}_{\text{bounary}}$ commute with each other. This means that the algebra of bulk operators maps to the endomorphisms of $\mc{O}_{\text{boundary}}$ viewed as a module for the algebra of charges. In symbols, we have an algebra homomorphism 
\begin{equation} 
\mc{O}_{\text{boundary}} \to \op{End}_{\oint{\mc{O}_{\text{boundary}}}} (\mc{O}_{\text{bounary}}). 
 \end{equation}
Since modules for $\oint \mc{O}_{\text{boundary}}$ are the same as modules for the vertex algebra, we see that  the algebra of bulk operators has a natural homomorphism to the endomorphisms of the vacuum module of the boundary vertex algebra. 

It is natural to expect that the same statement holds at the derived level. The derived version of the endomorphisms of the vacuum module is the self-$\Ext$'s of the vacuum module. By this argument, we find a homomorphism of algebras from the bulk operators to the self-$\Ext$'s of the boundary of the vacuum module of the boundary algebra.

Why do we conjecture that this map is an isomorphism? To understand this, it is fruitful to look at the analog of the statement we are making that holds for topologically twisted $N=(2,2)$ models in $2$ space-time dimensions.  In that setting, it is known \cite{Kapustin:2004df, Costello:2004ei} that the algebra of bulk operators is the Hochschild cohomology of the category of branes. Let us choose a generating object of the category of branes, whose algebra of boundary operators is $A_{\text{boundary}}$.  Then the algebra of bulk operators is the Hochschild cohomology of the algebra $A_{\text{boundary}}$.  

In both the $3d$ $N=4$ and $2d$ $N=(2,2)$ settings, we can define a module for the boundary algebra to be a way of changing the boundary algebra at a single point. Equivalently, a module in this sense is the end-point of a bulk line defect. In the $3d$ $N=4$ setting, these are ordinary modules for the boundary vertex algebra. In the $2d$ $N=(2,2)$ setting, these modules are bi-modules for the associative algebra $A_{\text{boundary}}$ of boundary operators.  In each situation, there is a special module in which the boundary algebra is unchanged, corresponding to the trivial line defect in the bulk.  In the $3d$ $N=4$ case, this special module is the vacuum module. In the $2d$ $N=(2,2)$ case, this special module is $A_{\text{boundary}}$ viewed as an $A_{\text{boundary}}$-bimodule.

Hochschild cohomology is the self-$\Ext$'s of $A_{\text{boundary}}$ taken in the category of $A_{\text{boundary}}$-bimodules.  This description makes clear the close analogy between self-$\Ext$'s of the vacuum module of the boundary VOA and Hochschild cohomology of the algebra of boundary operators.

In the $2d$ $N=(2,2)$ setting, we can only recover the algebra of bulk operators  from the algebra of boundary operators as long as the chosen boundary condition is ``big enough'', meaning it generates the category of boundary conditions. For example, if we are studying the $B$-twist of a two-dimensional $\sigma$-model on some Calabi-Yau manifold $X$, we will never learn about the entire algebra of bulk operators from a Dirichlet boundary condition in which the boundary fields map to a point $x$ in the target manifold $X$.  Instead, the Hochschild cohomology of the boundary algebra for this boundary condition will tell us about bulk operators in an infinitesimal neighourhood of this point $x$ in the target manifold $X$.

Similarly, in the three dimensional setting, we would not expect the algebra of boundary operators to always recover the algebra of bulk operators.  We conjecture that this is  true, however, as long as the theory flows to a CFT in the IR.   The conjecture can be shown to be false if we do not include this extra hypothesis.

To understand the need for this extra hypothesis, we note that boundary conditions which, after deformation, are compatible with the $SU(2)_H$-twist give rise to a complex submanifold of the Coulomb branch, and boundary conditions compatible with the $SU(2)_C$-twist give a complex submanifold of the Higgs branch.  In each case the submanifold is that associated to an ideal in the algebra of bulk local operators in the twisted theory, which is the algebra of holomorphic functions on the Coulomb or Higgs branch, depending on the twist. The ideal consists of those operators which become zero when brought to the boundary.

One can show that the submanifold associated to the deformed $(0,4)$ boundary condition is always isotropic (where we equip  the Higgs or Coulomb branch with the natural holomorphic symplectic structure).  Typically these submanifolds are not Lagrangian: Lagrangian submanifolds correspond to $(2,2)$ boundary conditions.

For general reasons, the self-$\Ext$'s of a given boundary condition can only know about the infinitesimal neighbourhood of the corresponding isotropic submanifold in the Higgs or Coulomb branch. If the theory is conformal, however, this is enough. In a conformal theory, the Higgs and Coulomb branch are conical, and for a reasonable boundary condition, the isotropic submanifold will be conical.  All the data of the Higgs and Coulomb branch is encoded in a neighourhood of the cone point, and so can in principle be detected by any boundary condition whose corresponding isotropic submanifold is conical. 

Let us describe a simple counter-example to our conjecture in the non-conformal case. Consider the free $U(1)$ gauge theory. The Coulomb branch in this case is $T^\ast \C^\times$, where $\C^\times$ is parametrized by the periodic scalar dual to the gauge field and $\C$ is parametrized by the scalar in the vector multiplet.  The fact that the scalar dual to the gauge field is periodic shows us that this theory is indeed not conformal.  From the point of view of the fundamental gauge field, this is a non-perturbative phenomenon that can be detected by monopole operators. 

Neumann boundary conditions for the gauge field are compatible (after deformation) with the $SU(2)_H$-twist,  and correspond to Dirichlet boundary conditions for the dual periodic scalar.  One can check that the scalar in the vector multiplet also  has Dirichlet boundary conditions. Boundary values of the bulk fields parameterize a submanifold of the Coulomb branch, which in this case is a point inside $\C \times \C^\times$.   

The self-$\Ext$'s of the boundary algebra can only probe an infinitesimal neighbourhood of this point in $\C \times \C^\times$, and one can indeed show that the self-$\Ext$'s are the algebra $\C[[z_1,z_2]]$ of formal series in two variables.  In particular, the self-$\Ext$'s can not tell us that the dual scalar is periodic.

\subsection{The free hypermultiplet in the $SU(2)_H$-twist }
Let us explain how to verify this conjecture in the case of a free hypermultiplet. 

If we perform the $SU(2)_H$-twist, the algebra of boundary operators is $\Sb$, the symplectic bosons.   We will change notation slightly, and write the symplectic bosons as $X_1,X_2$ instead of $X,Y$. We will view the category of modules for this vertex algebra as the category of modules for the algebra of charges. The algebra of charges is generated by 
\begin{align} 
X_{1,n} &= \oint z^{n} X_1(z) \d z \\
X_{2,n} &= \oint z^{n} X_2(z) \d z 
 \end{align}
with commutators
\begin{equation} 
[X_{1,n}, X_{2,m} ] = \delta_{n+m = -1}. 
 \end{equation}
The vacuum module is generated by a vector $\vac$ annihilated by $X_{i,n}$ for $n \ge 0$.  

Let us denote the algebra of charges by $\mc{A}$ and the vacuum module by $\mc{M}_{\vac}$.   The vacuum module admits a free resolution $\mc{A} [\eta_{i,n}]$, in which we have adjoined infinitely many odd variables $\eta_{i,n}$ to $\mc{A}$. The indices in $\eta_{i,n}$ run from $i = 1,2$ and $n \ge 0$.  The differential is
\begin{equation} 
\d = \sum \partial_{\eta_{i,n}} r_{X_{i,n}} 
 \end{equation}
 where $r_{X_{i,n}}$ indicates right multiplication with the generator $X_{i,n}$ of $\mc{M}_{\vac}$.  The odd variables $\eta_{i,n}$ are given cohomological degree $-1$.  

This differential makes $\mc{A}[\eta_{i,n}]$ into a differential graded left module for $\mc{A}$. The zeroth cohomology of this dg module is $\mc{M}_{\vac}$, and one can check that the other cohomology groups vanish.

The self-$\Ext$'s of $\mc{M}_{\vac}$ are the cohomology of the complex of maps of $\mc{A}$-modules from the free resolution $\mc{A}[\eta_{i,n}]$ to itself, or equivalently, from $\mc{A}[\eta_{i,n}]$ to $\mc{M}_{\vac}$.  This complex is $\mc{M}_{\vac}[\eta^\ast_{i,n}]$, where $\eta^\ast_{i,n}$ are odd variables dual to $\eta_{i,n}$, with differential
\begin{equation} 
\d = \sum_{i, n \ge 0} \eta^\ast_{i,n} X_{i,n}. 
 \end{equation}
Here the charges $X_{i,n}$ act in the usual way on the vacuum module. They commute with the odd variables $\eta^\ast_{i,n}$ and increase spin by $n + \tfrac{1}{2}$. 

The vacuum module $\mc{M}_{\vac}$ is freely generated from the vacuum vector by the lowering operators $X_{i,n}$, $n < 0$.  We can thus write the vacuum module as the polynomial algebra $\C[X_{i,n}, n < 0]$.  For $n \ge 0$, the charge $X_{i,n}$ acts as $\eps_{ij} \partial_{X_{j,-1-n}}$ where $\eps_{ij}$ is the alternating symbol.   After a relabelling of the odd variables $\eta^\ast_{i,n}$ by $\gamma_{i,-1-n} = \eps_{ij} \eta^\ast_{j,n}$ we find that the differential takes the form
=\begin{equation} 
\d = \sum_{i, n < 0} \gamma_{i,n} \partial_{X_{i,n}} 
 \end{equation}
acting on the polynomial algebra $\C[X_{i,n}]$.  This is simply the algebraic de Rham operator on the infinite-dimensional space with coordinates $X_{i,n}$ for $n < 0$.  Therefore the cohomology consists of $\C$ in degree $0$. 

This is the expected answer, because a free hypermultiplet has no Coulomb branch.
\subsection{The free hypermultiplet in the $SU(2)_C$-twist} 
Next, let us explain what happens for the $SU(2)_C$ twist. In this case, the boundary algebra is the algebra of fermionic currents, which we write as $x_1,x_2$ isntead of $x,y$. The associative algebra of charges is generated by 
\begin{equation} 
 x_{i,n} = \oint z^{n} x_{i} \d z   
 \end{equation}
 of spin $-n$ with commutators
\begin{equation} 
[x_{i,n}, x_{j,m}] = \eps_{ij} n \delta_{n+m = 0}.  
 \end{equation}
The vacuum module is generated by a vacuum vector annihilated by $x_{i,n}$ for $n \ge 0$.   Note that the $x_{i,0}$ are central. 

Following the analysis in the case of symplectic bosons, we find a free resolution of the vacuum module $\mc{M}_{\vac}$ by adjoining to the algebra $\mc{A}$ of charges an infinite number of generators $\phi_{i,n}$, for $i = 1,2$ and $n \ge 0$. In contrast to the case of symplectic bosons, these generators are bosonic, because the fundamental fields are fermionic.  The differential is $\sum x_{i,n} \partial_{\phi_{i,n}}$.  

The self-$\Ext$'s of the vacuum module then become $\mc{M}_{\vac}[\phi_{i,n}^\ast]$, where the ranges of the indices on $\phi_{i,n}^\ast$ are $i = 1,2$, $n \ge 0$.  The differential is
\begin{equation} 
 \d =  \sum_{n \ge 0} \phi_{i,n}^\ast x_{i,n}.  
 \end{equation}

We can identify $\mc{M}_{\vac}$ with the polynomial algebra on $x_{i,n}$ when $n < 0$. The operators $x_{i,n}$ for $n \ge 0$ become $\eps_{ij} n \partial_{x_{j,-n}}$.  We let $\sigma_{i,n} = \eps_{ij} \phi_{i,-n}^\ast$ for $n \le 0$.   We find that  the complex computing the self-$\Ext$'s is $\C[x_{i,n}, \sigma_{i,n}]$, where $x$ are odd variables and $\sigma$ are even. The odd generators $x_{i,n}$ have index $ n < 0$, and the even generators $\sigma_{i,n}$ have index $n \le 0$.  The differential is 
\begin{equation} 
\sum_{n < 0} \sigma_{i,n} n \partial_{x_{i,n}}. 
 \end{equation}

This is the tensor product of the de Rham complex on the infinite dimensional space with coordinates $\{x_{i,n} \mid n < 0\}$, with the polynomial algebra on $\sigma_{i,0}$.   After taking cohomology, the result is the polynomial algebra\footnote{Strictly speaking, we find the power-series algebra $\C[[\sigma_{i,0}]]$ as the self-$\Ext$'s. In this case, however, because of the $SU(2)$ symmetry rotating the $\sigma_{i,0}$, we can restrict to the subspace consisting of finite sums of elements which are in irreducible representations of $SU(2)$. We will typically not be very careful about the difference between polynomial and power series algebras.  } $\C[\sigma_{i,0}]$ on two variables.   

This is the desired answer, becauses the Higgs branch of a hypermultiplet is $\C^2$.

\subsection{A computation for a $U(1)$ gauge field}
As we will explain in detail shortly, if we perform an $SU(2)_H$-twist to a $3d$ $N=4$ gauge theory, then there is a deformable $(0,4)$ boundary condition with Neumann boundary conditions for all fields.  The boundary algebra is the BRST reduction of a system of symplectic bosons associated to the matter by the gauge group.  (In general we need to add extra boundary degrees of freedom to cancel an anomaly). 

If we start with a $U(1)$ pure gauge theory, then the boundary algebra is the BRST reduction of the trivial theory by $U(1)$.  This algebra has the $\b,\c$ ghosts which are fermionic and of spins $1$ and $0$.  The OPE is $\b \c \simeq 1/z$.  There is a very important subtlety, however: the $\c$ ghost by itself should not appear in the algebra, only its $z$-derivatives $\partial_z^k \c$ can appear.  This subtlety applies any time we introduce ghosts for gauge transformations in a compact group: we should only introduce ghosts for non-constant gauge transformations, and impose gauge invariance for the constant gauge transformations directly.

If we bear this subtlety in mind, we see that the algebra is generated by two fermionic fields $\b, \partial_z \c$ of spin $1$ with OPE $\b \partial_z \c \simeq z^{-2}$. This is the algebra of fermionic currents which we find when studying the $SU(2)_C$-twist of a free hypermultiplet.

This is a rather satisfying answer, because the $SU(2)_H$-twist of a $U(1)$ gauge theory is dual to the $SU(2)_C$ twist of a free hypermultiplet living in the cotangent bundle of $\C^\times$.  As we have seen, the Neumann boundary conditions for the gauge field become Dirichlet boundary conditions for free hypermultiplet. The algebra of operators with Dirichlet boundary conditions can not tell the difference between a periodic or non-periodic hypermultiplet, and so will be the algebra of fermionic currents.  In this way, we have verified that our boundary vertex algebras are compatible with the very simplest of dualities: a free $U(1)$ gauge field becoming a periodic hypermultiplet.  

Previously we found that the self-$\Ext$'s of the vacuum module of the hypermultiplet are $\C[[x_1,x_2]]$, the ring of formal series in two variables $x_1,x_2$.    As we have explained above, this is \emph{not} the Coulomb branch of the free $U(1)$ gauge theory. The full Coulomb branch is $T^\ast \C^\times$, and the self-$\Ext$'s of the vacuum module only recovers a small open subset of the Coulomb branch near the point where the dual periodic scalar is $1$ and the scalar in the  vector multiplet is $0$.

We have verified our conjecture for free hypermultiplets, and analyzed how it fails for a free $U(1)$ gauge theory.  In section \ref{subsection_detailed} we will show that the algebra of functions on the Coulomb branch for $U(1)$ gauge theory with one hyper is the self-$\Ext$'s of the vacuum module for the boundary VOA. We leave further checks of this conjecture to a separate publication \cite{VOAExt}. 

\subsection{Bulk lines and modules}
The physical theories also admit two classes of half-BPS line defects \cite{Assel:2015oxa}, which become topological line defects 
upon H- or C- twists. These preserve supercharges which have the same weight 
under the Cartan generators of rotations and $SU(2)_H$ or $SU(2)_C$:
either
\begin{equation}
Q^{+ \dot A}_+ \qquad \qquad Q^{- \dot A}_-
\end{equation}
or 
\begin{equation}
Q^{A \dot +}_+ \qquad \qquad Q^{A \dot -}_-
\end{equation}

Thus both types of line defects are compatible the holomorphic twist (along the $x^3$ direction)
and each the appropriate topological twist. These line defects can end on $(0,4)$ boundaries and at the 
endpoints one will find modules of the boundary VOAs.

Bulk line defects form a braided tensor category, with morphisms given by spaces of local operators 
joining line defects. We expect these morphisms and the whole braided category manifests itself as the corresponding
category of modules for the boundary VOAs and their $\Ext$ groups, though strictly speaking the setup only predicts the existence of a 
functor from the bulk braided tensor category to the category of boundary modules.   

Let us explain how this should work for the free hypermultiplet, when we perform the $SU(2)_C$-twist. The bulk theory becomes Rozansky-Witten theory on $\C^2$, and it is expected that the category of line defects is the category of coherent sheaves on $\C^2$, or equivalently, the category of modules over the polynomial algebra $\C[z_1,z_2]$.  This is equivalent\footnote{A little care is needed in this equivalence: we should either restrict to the category of modules which are compatible with the grading which assigns to the $z_i$ charge $1$, or else look at the category of modules over the ring of power series $\C[[z_1,z_2]]$. We will not belabour this technical point.} to the category of modules over the exterior algebra $\C[x_1,x_2]$ generated by two odd variables.  This equivalence is the basic example of Koszul duality. 

Under this equivalence, a coherent sheaf $F$ gets sent to $\Ext^\ast(\C_0, F)$ where $\C_0$ is the skyscraper sheaf at the origin in $\C^2$.  This is a module for $\Ext^\ast(\C_0,\C_0)$, which is the exterior algebra on two generators.  

The exterior algebra on two generators can, in turn, be viewed as the universal enveloping algebra of the Abelian fermionic Lie algebra $\Pi \C^2$. This Lie algebra has an invariant symmetric pairing, given by the symplectic form on $\C^2$.  The algebra of fermionic currents is the Kac-Moody algebra built from this Lie algebra, at any non-zero level (all non-zero levels can be related by rescaling the generators). 

Modules for the exterior algebra $\C[x_1,x_2]$ are then the same as modules for the Lie algebra $\Pi \C^2$, i.e.\ super-vector spaces with two commuting odd symmetries.  Given any such module $M$, we can build a Weyl module $W(M)$ for the algebra of fermionic currents.  This Weyl module is generated by vectors $m \in M$, annihilated by $x_{i,n}$ for $n > 0$, and which transform under $x_{i,0}$ according to the action of $\Pi \C^2$ on $M$.   

We expect that the braiding of lines operators in the bulk theory of a free hypermultiplet is the braiding of the corresponding Weyl module for the fermionic current algebra.  

Explicit formulas can be obtained by considering the Knizhnik-Zamolodchikov connection for the Abelian fermionic Lie algebra $\Pi \C^2$.  Given representations $M_1,\dots, M_n$ of $\Pi \C^2$, in which the two elements of $\Pi \C^2$ act by matrices $x_{i}^r$ ($i = 1,2$, $r = 1,\dots,n$) we can define a connection on the trivial vector bundle on $\C^n$ with fibre $M_1 \otimes \dots \otimes M_n$ by the one-form
\begin{equation} 
	\sum_{\substack{r\neq s}} \frac{x_i^r x_j^s \eps_{ij} }{z_r - z_s} \d z_r .  
\end{equation}

We have seen that the boundary algebra for a pure $U(1)$ gauge theory is also the algebra of fermionic currents. In this example, we do not expect an equivalence of categories between bulk lines and boundary modules, only a functor.  It would be interesting to analyze the modules for the fermionic current algebra coming from line operators of the $U(1)$ gauge theory. 

\subsection{Topological boundary conditions}
The $(2,2)$ boundary conditions, which become topological upon twisting, produce particularly nice states in the TFT Hilbert space, 
which have good behaviour under mapping class group transformations. They should correspondingly map to special 
conformal blocks for the boundary VOA. 

Indeed, a slab geometry with a deformed $(0,4)$ boundary condition at one end and a $(2,2)$ boundary condition at the other hand 
should give a two-dimensional system, whose operator algebra includes both the original boundary VOA and any modules attached to 
lines which can end at the $(2,2)$ boundary condition. This is a non-trivial {\it extension} of the original VOA and the 
special conformal blocks must be those for which the fields of the extended VOA are single-valued. 

Furthermore, boundary local operators at $(2,2)$ boundary conditions give interesting modules for the algebra of bulk local operators
\cite{Bullimore:2016nji}, which should be reflected in the properties of this VOA extension. This opens up the possibility of a direct 
connection between the properties of boundary VOAs and the aspects of Symplectic Duality associates to $(2,2)$ boundary conditions. 

Again, we leave a detailed analysis to future work. 

\section{The H-twist of standard $\CN=4$ gauge theories} \label{sec:H}
We consider here standard $\CN=4$ gauge theories, defined by vectormultiplets in a gauge group $G$ and 
hypermultiplets in a symplectic representation $M$ of $G$. 

Based on the elementary examples and calculations in the holomorphically twisted theory, we expect the following class of 
$(0,4)$ boundary conditions to be compatible with the deformation to the H-twisted theory: 
\begin{itemize}
\item Neumann boundary conditions for the vectormultiplets.
\item Neumann boundary conditions for the hypermultiplets.
\item Extra boundary degrees of freedom, in the form of an holomorphic CFT $A_{2d}$ with a $G$ current algebra 
coupled to the bulk gauge fields at the boundary. 
\end{itemize}

Notice that both the boundary conditions for the hypermultiplets and for the vectormultiplets 
introduce potential gauge anomalies. The extra boundary degrees of freedom can be used to cancel that. 

The half index for such a boundary condition takes precisely the form of the character for a $\mathfrak{g}$-BRST reduction 
of the product VOA
\begin{equation}
\Sb^M \times A_{2d}
\end{equation}
of symplectic bosons valued in $M$ together with the auxiliary boundary VOA. Boundary anomaly cancellation 
precisely matches the requirement that the total level of the $G$ current algebra if $-2h$, the valued required for the BRST reduction. 

Concretely, that means considering the BRST complex \cite{1990NuPhB.329..649K}
\begin{equation}
\left[ \Sb^M \times A_{2d} \times \mathrm{bc}^{\mathfrak{g}}, Q^\mathfrak{g}_{\mathrm{BRST}} = \oint c \left(J_{\Sb^M \times A_{2d}} + J_{ \mathrm{bc}^{\mathfrak{g}}}\right) \right]
\end{equation}
This BRST reduction is our candidate boundary VOA. 

The following observations are in order:
\begin{itemize}
\item For classic gauge groups, $A_{2d}$ can usually be taken to be some collection of chiral free fermions. 
For conciseness, we can denote the $G$-BRST reduction of the product of 
symplectic bosons valued in a representation $M$ and chiral fermions valued in a representation $R$ as:
\begin{equation}
\mathfrak{A}_H[G,M,R] := \{ \mathrm{Sb}[M] \times \mathrm{Ff}[R] \times \mathrm{bc}^{\mathfrak{g}}, Q^\mathfrak{g}_{\mathrm{BRST}} \}
\end{equation}
\item The association of a VOA to the H-twisted 3d gauge theory was proposed first in \cite{Gaiotto:2016wcv}, with $A_{2d}$ consisting of a 
chiral WZW model of the appropriate level. That choice is not ideal, as chiral WZW models are only {\it relative} theories. 
We will revisit and improve that construction in our examples. 
\item The same type of BRST reduction, without auxiliary degrees of freedom, appears in the construction 
of chiral algebras associated to 4d $\CN=2$ gauge theories. Compactification of a 4d $\CN=2$ gauge theory 
on a cigar geometry yields precisely our $(0,4)$ boundary condition. The further deformation we introduce to 
go to the topologically twisted 3d theory should be analogous to a 
twisted Nekrasov deformation in the 4d $\CN=2$ gauge theory, employing the $U(1)_r$ R-symmetry group instead of the 
$SU(2)_R$ group employed in the traditional Nekrasov deformation. The fact that the 4d $\CN=2$ theory is super-conformal implies that the BRST reduction on the boundary of the 3d theory is anomaly free, without the need to introduce extra degrees of freedom.  
		
It would be interesting to fully explore the 
properties of such a twisted Nekrasov deformation and the relation with the super-conformal twist used in the definition of the 
chiral algebras associated to 4d $\CN=2$ gauge theories.
\end{itemize}

\subsection{Symmetries of H-twistable boundaries}
The boundary conditions compatible with the H-twist preserve the global flavor symmetry $G_H$ 
which acts on the hypermultiplets. 

The bulk gauge theory also has a global flavor symmetry 
$G_C$ which acts on the Coulomb branch. Only the Cartan sub-algebra $U(1)^r_C$ 
is visible in the UV, as topological symmetries whose currents are the gauge field strength.

Neumann boundary conditions for the gauge field naively break the topological symmetry 
of the gauge theory, as the inflow of charge into the boundary equals the gauge field strength 
at the boundary, which is unconstrained. A topological symmetry can be restored by combining the bulk symmetry  
with a boundary symmetry $U(1)_{2d}$ which has exactly one unit of mixed anomaly with the corresponding gauge symmetry: 
the divergence of the 2d current equals the gauge field strength at the boundary, which is the 
inflow of the 3d charge. 

As the Neumann boundary conditions for Abelian gauge fields generically do require 
extra boundary Fermi multiplets, we can typically use the 2d symmetries rotating these Fermi multiplets 
in order to restore the bulk $U(1)^r_C$ symmetry algebra. With a bit of luck, the resulting boundary condition may preserve the whole $G_C$
which appears in the IR. 

The boundary Fermi multiplets may transform under a further symmetry group $G_{2d}$ commuting with the gauge 
group, modulo the symmetries we absorbed in $U(1)^r_C$. 

\subsection{Conformal blocks and $\Ext$ groups}
It is interesting to ask which properties of the $\mathfrak{A}_H[G,M,R]$ would follow directly from its definition. 
For example, can be compute some (or all) conformal blocks for $\mathfrak{A}_H[G,M,R]$ that way? 
Clearly, we can attempt to realize such conformal blocks as correlation functions of BRST-closed operators in 
$\mathrm{Sb}[M] \times \mathrm{Ff}[R] \times \mathrm{bc}[\mathfrak{g}]$. 

The $\mathfrak{g}$-ghosts have zeromodes on a general Riemann surface $C$. Analogously to what is done in string theory, 
we can compensate for these zeromodes with $b$-ghost insertions. These insertions make correlation functions into 
top forms in the space $\Bun_G(C)$ of $G$-bundles on the Riemann surface, which should be formally integrated over $\Bun_G$. 

Thus we can realize conformal blocks for $\mathfrak{A}_H[G,M,R]$ by taking conformal blocks for $\mathrm{Sb}[M] \times \mathrm{Ff}[R]$,
seen as a D-module on $\Bun_G(C)$, and taking de Rahm cohomology over $\Bun_G$. Note that $\mathrm{Ff}[R]$ 
(or more generally any well-defined 2d degrees of freedom) have one-dimensional spaces of conformal blocks.  

If we apply this idea to the $\Ext$ groups, we need to do calculations with bundles over the raviolo. 
We expect these calculations to directly reproduce the definitions in \cite{Nakajima:2015txa,Braverman:2016wma}. 
We will include a more detailed argument in upcoming work \cite{VOAExt}.

\subsection{$U(1)$ gauge theory with one flavor}
This gauge theory is mirror to a free twisted hypermultiplet valued in $\C^2$ \cite{Kapustin:1999ha}. 
\footnote{The target is really Taub-NUT, but in the IR it flows to a flat target}
The global symmetry of the twisted hypermultiplet is identified with the topological symmetry $U(1)_t$
of the gauge theory, whose current is the gauge field strength. 

The H-twist compatible boundary conditions require a single Fermi multiplet of gauge charge $1$ at the boundary,
for gauge anomaly cancellation \cite{Dimofte:2017tpi}. We can take $U(1)_t$ to act on the Fermi multiplet with charge $1$ in order to define an unbroken symmetry. No extra boundary symmetries remain. Thus the overall symmetry of the system is $U(1)_t$. 

This is compatible with the mirror description of the boundary condition to be the 
basic C-twist compatible boundary condition for the free twisted hypermultiplet, 
i.e. Dirichlet boundary conditions for twisted hypermultiplet scalars. 

Physically, this is sensible: the twisted hypermultiplets are realized as monopole operators 
in the bulk gauge theory. Monopole operators brought to the boundary disappear, leaving behind 
gauge-invariant local operators with the same $U(1)_t$ charge. The simplest such operators are the product of a 
(chiral) boundary fermion and the boundary value of a hypermultiplet scalar. This process should be mirror to a twisted 
hypermultiplet going to a Dirichlet boundary: the scalars vanish and the fermionic chiral components survive. 

The C-twist compatible boundary condition preserves $U(1)_t$ and the boundary symmetry $U(1)_{2d}$
from the bulk gauge symmetry. The $U(1)_{2d}$ has a `t Hooft anomaly because of the hypermultiplet boundary conditions. 
The $U(1)_t$ and $U(1)_{2d}$ symmetries have a mixed `t Hooft anomaly at the boundary. $U(1)_t$ has no boundary  `t Hooft anomaly.
Bulk monopoles brought to the boundary will now map to boundary monopoles. 

It is not hard to propose a candidate mirror: a C-twist compatible Neumann boundary condition for the twisted hypermultiplet, 
enriched by an extra free Fermi multiplet at the boundary, charged under $U(1)_{2d}$ and $U(1)_t$, which cancels the $U(1)_t$
anomaly induced by the twisted hypermultiplet boundary condition. 

Notice that on both sides of the mirror symmetry relations we either find (twisted) hypers with Dirichlet b.c. or 
(twisted) hypers with Neumann b.c. paired up with a Fermi multiplet of the same charge.

Closely related mirror symmetry relations for boundary conditions were studied recently in \cite{Dimofte:2017tpi} 
and tested at the level of the index. 

We can readily test the mirror symmetry at the level of the boundary VOA. On the H-twisted side, the algebra $\mathfrak{A}_H[U(1),\C^2,\C^2]$ 
is built as the $\mathfrak{u}(1)$ BRST reduction of the product 
\begin{equation}
\Sb \times \Ff
\end{equation}
of a symplectic boson pair and a free complex fermion VOAs. 

Denote the symplectic bosons as $X$,$Y$ and complex fermions as $\chi$, $\psi$,
with OPE
\begin{equation}
X(z) Y(w) \sim \frac{1}{z-w} \qquad \qquad \chi(z) \psi(w) \sim \frac{1}{z-w}
\end{equation}
The BRST charge involves the total $\mathfrak{u}(1)$ current $J_{\mathrm{tot}} = X Y - \chi \psi$ which gives charge $1$ to 
$X$ and $\chi$ and $-1$ to $Y$ and $\psi$. 

Bilinears $XY$, $\chi \psi$, $\chi Y$ and $X \psi$ of charge $0$ for $J_{\mathrm{tot}}$ form a set of  $\mathfrak{u}(1|1)_{-1}$ Kac-Moody currents. 
\footnote{The full set of bilinears actually forms a set of $\mathfrak{osp}(2|2)_1$ currents, but we will not need that.}
Indeed, the whole charge $0$ sector of the algebra can be identified with the $\mathfrak{u}(1|1)_{-1}$ Kac-Moody algebra,
say by matching characters. The other charge sectors transforms as interesting modules for $\mathfrak{u}(1|1)_{-1}$, 
but they will drop out of the BRST reduction. \footnote{The whole $\Sb \times \Ff$ can be seen as a sort of ``WZW model'' 
associated to $\mathfrak{u}(1|1)_{-1}$, in a sense that we will explain better in Section \ref{sec:C}. }

The $\mathfrak{u}(1)$ BRST reduction thus acts directly on the $\mathfrak{u}(1|1)_{-1}$ Kac-Moody algebra. As we will see in detail shortly,   it  reduces it to a $\mathfrak{psu}(1|1)$ Kac-Moody algebra, generated by two BRST-closed fermionic currents $x= X \psi$ and $y= Y \chi$ 
with OPE 
\begin{equation}
x(z) y(w) \sim \frac{1}{(z-w)^2}
\end{equation}
This is the same as the VOA for the conjectural mirror: a Dirichlet boundary conditions for a free twisted hypermultiplet!

This statement can be easily checked at the level of half indices/characters for the VOA (and we will give an explicit derivation at the level of the VOA itself in the next section).
The character for the BRST reduction reads
\begin{equation}
(q;q)^2_\infty \oint \frac{dz}{2 \pi i z} \frac{(y z q^{\frac12};q)_\infty(y^{-1} z^{-1} q^{\frac12};q)_\infty}{(z q^{\frac12};q)_\infty(z^{-1} q^{\frac12};q)_\infty} = (y q;q)_\infty(y^{-1} q;q)_\infty
\end{equation}
The equality can be proven with the tools in \cite{Dimofte:2017tpi}.

\subsection{A detailed verification of the duality}
\label{subsection_detailed}
We have sketched above that the BRST reduction of two symplectic bosons with a pair of complex fermions, under the $U(1)$ action with current $X Y - \chi \psi$, should be the fermionic current algebra. This represents the duality between $U(1)$ with one hyper and one free hyper, at the level of boundary algebras.  In this section we will verify this in detail, by explicitly calculating the BRST cohomology.  

At a first pass, the BRST complex is obtained by adjoining to the symplectic boson and free fermion system $\Sb \times \Ff$ a $\b$-ghost and a $\c$-ghost, of ghost numbers $-1,1$ and spins $1,0$.  The BRST operator is defined by
\begin{align} 
	Q \b &= J_{\mathrm{tot}} = XY - \chi \psi \\
	Q \alpha &= \sum \frac{1}{n!}\partial_z^n \c \oint z^n J_{\mathrm{tot}}(z) \alpha(0)  \d z, \label{eqn_brst}
\end{align}
where $\alpha$ is any local operator in the $\Sb \times \Ff$ system.  
This is not quite correct, however, as we should not include the $\c$-ghost itself in the BRST complex, only its derivatives.  The constant $\c$-ghost enforces gauge invariance for constant gauge transformations. The correct definition of BRST reduction is defined by adjoing to the charge $0$ sector of $\Sb \times \Ff$ a pair of fermionic currents denoted $\b$ and $\partial_z \c$, with BRST operator defined by equation \ref{eqn_brst}.

To calculate this, we first need to describe more carefully the charge $0$ sector $(\Sb \times \Ff)^0$ of $\Sb \times \Ff$. We stated above that this algebra should be a quotient of the $\mathfrak{u}(1 \mid 1)_{-1}$, where the generators of the Kac-Moody algebra map to the charge $0$ bilinears $XY$, $\chi \psi$, $X \psi$, $\chi Y$.  It is not completely obvious, however, that the algebra of charge $0$ operators is generated by these bilinears. 

Indeed, the charge $0$ sector of just the symplectic bosons is \emph{not} generated by the bilinear $XY$.  There are three operators of spin $2$ in the charge $0$ sector of the symplectic bosons, namely $X\partial_z Y$, $\partial_z X Y$, and $X^2 Y^2$; whereas there are only two operators of spin $2$ in the $U(1)$ current algebra.  

For the charge $0$ sector of $\Sb \times \Ff$, this problem does not arise: the algebra of charge $0$ operators is generated by the four $\mathfrak{u}(1 \mid 1)_{-1}$ currents. To see this, we first note that the charge $0$ algebra is generated by the operators $X \partial_z^n Y$, $\chi \partial_z^n \psi$, $X \partial_z^n \psi$, $\chi \partial_z^n Y$.    We need to show that these operators can be obtained as iterated OPEs of the four bilinears which don't have any derivatives.  

Suppose, by induction, that all charge $0$ bilinears with $n-1$ derivatives are in the subalgebra generated by the $\mathfrak{u}(1 \mid 1)_{-1}$ currents. We will show that the charge $0$ bilinears with $n$ derivatives are also in this subalgebra. To see this, we note that 
\begin{align*} 
	\chi(0) \partial_z^n \psi (0) =& \oint \chi (0) \psi(0) \chi(z) \partial_z^{n-1} \psi(z) z^{-1} \d z \\	
X(0) \partial_z^n \psi (0) =& \oint X (0) \psi(0) \chi(z) \partial_z^{n-1} \psi(z) z^{-1} \d z \\ 
& + \oint X(0) \partial_z^{n-1} \psi(0) \chi(z) \psi(z) z^{-1} \d z \\
Y(0) \partial_z^n \chi(0) =& \oint Y(0) \chi(0) \psi(z) \partial_z^{n-1} \chi(z) z^{-1} \d z \\ 
& + \oint Y(0) \partial_z^{n-1} \chi(0) \psi(z) \chi(z) z^{-1} \d z \\
X(0) \partial_z^n Y(0) = & \oint X(0) \psi(0) \chi (z) \partial_z^{n-1} Y(z) z^{-1} \d z \\ 
& - \oint X(0) \partial_z^{n-1} Y(0) \psi(z) \chi(z) z^{-1} \d z.
\end{align*}
Each line expresses one of the bilinears with $n$ derivatives in terms of the non-singular term in the OPE between bilinears with $n-1$ and fewer derivatives.  

This completes the proof that the charge $0$ sector $(\Sb \times \Ff)^0$ is a quotient of $\mathfrak{u}(1 \mid 1)_{-1}$. 

We let
\begin{align} 
J_{\mathrm{tot}} &= XY - \chi \psi \\
\til{J} &= X Y + \chi \psi .
 \end{align}
 Note that $J_{\mathrm{tot}} \til{J} \simeq z^{-2}$. Therefore these operators together form the currents for $\mathfrak{u}(1)_1 \times \mathfrak{u}(1)_1 $.   We will decompose $(\Sb \times \Ff)^0$ as a module over the currents given by $J_{\mathrm{tot}}$, $\til{J}$. 

We let $J_n = \oint J_{\mathrm{tot}} z^{n} \d z$, and $\til{J}_n = \oint \til{J} z^n \d z$. These are operators acting on the the vacuum module for $(\Sb \times \Ff)^0$,  where $J_n, \til{J}_n$ for $n < 0$ are raising operators and $J_n, \til{J}_n$ for $n > 0$ are lowering operators.  

Let $(\Sb \times \Ff)^0_0$ denote the subspace of highest-weight vectors, that is, the elements of the vacuum module of $(\Sb \times \Ff)^0$ killed by all the lowering operators $J_n, \til{J}_n$ for $n > 0$. Then basic facts about the representation theory of the $\mathfrak{u}(1)$ current algebra tells us that the vacuum module for $(\Sb \times \Ff)^0$ is freely generated from $(\Sb \times \Ff)^0_0$ by an application of the lowering operators $J_n, \til{J}_n$ for $n < 0$.  That is,
\begin{equation} 
(\Sb \times \Ff)^0 = (\Sb \times \Ff)^0_0[J_{-1}, J_{-2}, \dots, \til{J}_{-1},\til{J}_{-2},\dots] . 
 \end{equation}

Next, we need to compute the BRST cohomology.   Looking at equation \eqref{eqn_brst}, we see that the BRST operator on the charge $0$ operators, with the $\b$ and $\partial_z \c$ ghosts adjoined, takes the form
 \begin{align} 
 Q \b &= J_{\mathrm{tot}} \\
 Q \til{J} &= \partial_z \c. 
  \end{align}
The vacuum module of the BRST reduction can be written
\begin{align} 
(\Sb \times \Ff)_{BRST} &=  (\Sb \times \Ff)^0[\b_{-1}, \b_{-2},\dots,(\partial_z\c)_{-1}, (\partial_z \c)_{-2}, \dots ]\\
&= (\Sb \times \Ff)^0_0 [J_{-1}, J_{-2}, \dots, \til{J}_{-1},\til{J}_{-2},\dots, \b_{-1}, \b_{-2}, \dots, (\partial_z \c)_{-1}, (\partial_z \c)_{-2}, \dots ] . 
 \end{align}
The BRST operator transfroms $\b_k$ into $J_k$ and $\til{J}_k$ into $(\partial_z \c)_k$.  The BRST operator is trivial on the subspace $(\Sb \times \Ff)^0_0$ of highest weight vectors in the charge $0$ sector of $\Sb \times \Ff$.  

From this it follows that the cohomology of $(\Sb \times \Ff)_{BRST}$ is concentrated in ghost number $0$ and is isomorphic to $(\Sb \times \Ff)^0_0$, the space of highest-weight vectors in the charge $0$ sector of $\Sb \times \Ff$.  Because the charge $0$ algebra is generated by the two bosonic currents $J_{\mathrm{tot}}$, $\til{J}$ and the two fermionic currents
\begin{align} 
x &= X \psi \\
y &= \chi Y 
\end{align}
we find that the space $(\Sb \times \Ff)^0_0$ can be generated from the vacuum by the fermionic currents $x,y$.   Since the BRST cohomology of the vacuum module is isomorphic to $(\Sb \times \Ff)^0_0$, we deduce that the BRST cohomology must be \emph{some} quotient of the algebra $\Fc$ of fermionic currents.

Finally, we note that a simple representation theory argument tells us that $\Fc$ does not admit any non-trivial quotients.  This completes the argument that the BRST cohomology is isomorphic to the algebra $\Fc$ of fermionic currents. 

We have gone through this example in such great detail because it provides the first non-trivial example of the main conjectures of this paper. The BRST quotient $(\Sb \times \Ff)_{BRST}$ is the algebra of boundary operators for $U(1)$ with one hypermultiplet.  We have found that it is equivalent to the algebra of boundary operators for one free hyper, which is the dual theory.  

We have already shown that the self-Ext's of the vacuum module of the fermionic current algebra is the algebra of functions on $\C^2$, which is the Higgs branch of one free hyper. The fact that the boundary vertex algebras are compatible with the duality tells us that the self-Ext's of the vacuum module for $(\Sb \times \Ff)_{BRST}$ is the same space, which is the Coulomb branch of $U(1)$ with one hyper. This is the first non-trivial check of our proposal for describing moduli of vacua in terms of boundary vertex algebras. 

One aspect of this description of the Coulomb branch is somewhat remarkable. The boundary VOA for $U(1)$ with one hyper was described entirely in perturbative terms. All boundary operators are functions of the fundamental fields, and the OPEs and BRST operator be derived from explicit semi-classical computations (see \cite{Companion} for more details).  Even so, the boundary VOA contains enough information to recover the monopole operators in the bulk, which are non-perturbative objects. 

\section{More elaborate examples of H-twist VOAs}
In this section we study a sequence of examples of increasing complexity. 

\subsection{$U(1)$ gauge theory with $N$ flavors}
The algebra $\mathfrak{A}_H[U(1),\C^{2N},\C^{2N}]$ is built as a $U(1)$ BRST coset of 
the product of $N$ sets of symplectic bosons $X^a$,$Y_a$ and complex fermions $\chi^i$, $\psi_i$
with OPE
\begin{equation}
X^a(z) Y_b(w) \sim \frac{\delta^a_b}{z-w} \qquad \qquad \chi^i(z) \psi_j(w) \sim \frac{\delta^i_j}{z-w}
\end{equation}
The BRST charge involves the total level $0$ $U(1)$ current $X^a Y_a + \chi^i \psi_i$ which gives charge $1$ to 
$X^a$ and $\chi^i$ and $-1$ to $Y_a$ and $\psi_i$. 

We will denote the charge $0$ sector of the VOA as $\mathrm{u}(N|N)_{1}$, as we expect it to be generated by 
$\mathrm{u}(N|N)_{1}$ currents defined as bilinears $X^a Y_b$, $X^a \psi_i$, $Y_a \chi^i$, $\psi_i \chi^j$. 
The $\mathrm{u}(N|N)_{1}$ subalgebra is clearly {\it not} the same as a 
$\mathfrak{u}(N|N)_{1}$ Kac-Moody sub-algebra. For example, the fermionic bilinears form an 
$\mathrm{u}(N)_{1}$ current algebra which includes an $\mathrm{su}(N)_{1}$ WZW simple quotient of
$\mathfrak{su}(N)_{1}$ Kac-Moody. 

The $U(1)$ BRST coset removes two of the currents, leaving behind a vertex algebra 
which contains an $\mathrm{psu}(N|N)_{1}$ current algebra. Again, we expect the vertex algebra to 
be generated by the $\mathrm{psu}(N|N)_{1}$ currents and to be some quotient of the $\mathfrak{psu}(N|N)_{1}$ Kac-Moody algebra.

For general $N$, the $\mathrm{psu}(N|N)_{1}$ VOA has an $U(1)_C$ outer automorphism acting on the two blocks of fermionic generators
with charges $\pm 1$. We will see that for $N=2$ this symmetry group is enhanced. 

For some values of $N$, we can also look at non-canonical choices of fermion representations. 
For example, we can consider $\mathfrak{A}_H[U(1),\C^{8},\C^{2}(2)]$, 
involving a single set of complex fermions of charge $2$. 

Typical operators in $\mathfrak{A}_H[U(1),\C^{8},\C^{2}(2)]$ are the $SU(4)_{-1}$ currents 
$X^a Y_b$ and the fermionic generators $X^a X^b \psi$ and $Y_a Y_b \chi$ of dimension $3/2$. 

\subsubsection{The $T[SU(2)]$ theory.}
The case $N=2$ is special because the corresponding gauge theory is expected to have a low-energy enhancement $U(1)_C \to SU(2)_C$.
Indeed, this is the $T[SU(2)]$ theory which plays a crucial role in S-duality for four-dimensional $SU(2)$ gauge theory \cite{Gaiotto:2016aa}.
The symmetry enhancement is crucial for that role and necessary for Geometric Langlands applications \cite{Gaiotto:2016wcv}.

Looking at the boundary VOA we built for the H-twisted theory, we see that the two blocks of fermionic generators have the same quantum numbers under the 
$\mathrm{su}(2)_1 \times \mathfrak{su}(2)_{-1}$ bosonic subalgebra. The $\mathrm{psu}(2|2)_{1}$ algebra has an $SU(2)_C$ outer automorphism 
and thus enjoys the full IR symmetry enhancement of the bulk theory!

Index calculations show a remarkable structure for $\mathfrak{A}_H[U(1),\C^{4},\C^{4}]$. The central charge of the VOA is $-2$ and coincides with the 
central charge of $\mathrm{su}(2)_1 \times \mathfrak{su}(2)_{-1}$, suggesting that the VOA may be a conformal extension of that 
current sub-algebra. Indeed, the character decomposes as 
\begin{equation}
\chi_{\mathrm{psu}(2|2)_{1}} = \sum_{j=0}^\infty \chi_{SU(2)_C}^{(j)}\chi_{\mathrm{su}(2)_{-1}}^{(j)}\chi_{\mathrm{su}(2)_{1}}^{(j \,\mathrm{mod} \,2)}
\end{equation}

We expect that conformal blocks for $\mathrm{psu}(2|2)_{1}$ should play the role of a kernel for the $SU(2)$ Geometric Langlands
when coupled both to $SU(2)$ flat connections through the $SU(2)_C$ outer automorphism and to $SU(2)$ bundles through the $\mathfrak{su}(2)_{-1}$ 
current algebra. \footnote{There are important subtleties to consider here concerning the global form of the gauge group, 
which will be discussed in a separate publication \cite{FG}. In short, $\mathrm{psu}(2|2)_{1}$ can be coupled to $SO(3)$ connections/bundles 
by coupling them to the $\mathrm{su}(2)_1$ currents as well.}

The existence of an algebraic coupling to $SU(2)$ flat connections is tied to the existence of a deformation/central extension of 
the $\mathrm{psu}(2|2)_{1}$ OPE involving coupling to a background holomorphic connection \cite{Gaiotto:2016wcv}.
In turn, this is an infinitesimal version of a more general deformation $\mathrm{psu}(2|2)_{1} \to \mathrm{d}(2,1,-\Psi)_{1}$
to a vertex algebra which appears at certain junctions of boundary conditions in GL-twisted ${\cal N}=4$ SYM \cite{Gaiotto:2017euk}.
and is associated to quantum Geometric Langlands duality. 

The coincidence of our boundary VOA with the $\Psi \to \infty$ limit of $\mathrm{d}(2,1,-\Psi)_{1}$ is quite remarkable, as the two 
VOAs are obtained by very different means. The coincidence will become somewhat less surprising once we look at the mirror construction 
of the C-twist boundary VOA, which can be continuously connected to the four-dimensional construction. 

The naive VOA proposed in \cite{Gaiotto:2016wcv} can be identified with the
\begin{equation}
V_{\mathrm{old}} = \frac{\mathrm{psu}(2|2)_1}{\mathrm{su}(2)_1}
\end{equation}
which strips off the $\mathrm{su}(2)_1$ fermion bilinears in the BRST complex leaving behind the $\mathrm{u}(1)_2$ lattice vertex algebra.
Conversely, $V[SU(2)]$ can be interpreted as an extension of $V_{\mathrm{old}} \times SU(2)_1$. 

\subsection{$SU(2)$ gauge theory with $N \geq 4$ flavors}
This gauge theory has $SO(2N)_H$ global symmetry. The gauge anomaly is 
$4-N$, which we cancel with $N-4$ doublets of Fermi multiplets. 

Thus the algebra is 
$\mathfrak{A}_H[SU(2),\C^{4N},\C^{4N-16}]$, defined as the $\mathfrak{su}(2)$ BRST quotient 
of the symplectic bosons $Z_\alpha^a$ and fermions $\zeta_\alpha^i$, $\alpha$ being the doublet index.

Simple gauge-invariant operators made as bilinears of symplectic bosons and fermions 
generate an $\mathrm{su}(2N|2N-8)_{-2}$ current algebra. That algebra has a Sugawara stress tensor 
of central charge matching the central charge of $\mathfrak{A}_H[SU(2),\C^{4N},\C^{4N-16}]$. 
It is not unreasonable to conjecture that $\mathfrak{A}_H[SU(2),\C^{4N},\C^{4N-16}]$ coincides 
with $\mathrm{su}(2N|2N-8)_{-2}$, to be thought of as some quotient of the $\mathfrak{su}(2N|2N-8)_{-2}$
Kac-Moody algebra. 

\subsection{$U(2)$ gauge theory with $N \geq 4$ flavors}
This gauge theory has $U(N)_H \times U(1)_C$ global symmetry. 

The $SU(2)$ gauge anomaly is $4-N$, which we cancel with $N-4$ doublets of Fermi multiplets. 
That leaves an anomaly for the diagonal $U(1)$ subgroup in $U(2)$. If we normalize 
that in such a way that a fundamental representation has charge $1/2$, then the residual anomaly is $-2$.
In order to cancel it, we add two more Fermi multiplets which are charged only under the diagonal $U(1)$. 

The overall symmetry algebra is thus $SU(N)_H \times U(1)_C\times SU(N-4)_{2d} \times SU(2)_{2d} \times U(1)_{2d}$.

The corresponding boundary VOA is $\mathfrak{A}_H[U(2),\C^{4N},\C^{4N-16} \oplus \C^4(2)]$, defined as the $\mathfrak{u}(2)$ BRST quotient 
of the symplectic bosons $X_\alpha^a$, $Y^\alpha_a$, fermions $\chi_\alpha^i$, $\psi^\alpha_i$, $\alpha$ being the doublet index,
and extra fermions $\tilde \psi^n$, $\tilde \chi_n$ in the determinant representations of $U(2)$. 

Simple operators made as $U(2)$-invariant bilinears of symplectic bosons $X$, $Y$ and fermions $\psi$, $\chi$
generate an $\mathrm{u}(N|N-4)_{-2}$ current algebra. The $\mathfrak{u}(1)$ part of the BRST coset will reduce that to 
$\mathrm{su}(N|N-4)_{-2}$ and remove the $\tilde \chi_n \tilde \psi^n$ current. The bilinears $\tilde \chi_n \tilde \psi^m$ give an 
$\mathrm{su}(2)_1$ WZW current algebra. The $\mathrm{su}(N|N-4)_{-2}\times \mathrm{su}(2)_1$ currents are associated to the 
$SU(N)_H \times SU(N-4)_{2d} \times SU(2)_{2d} \times U(1)_{2d}$ symmetries of the system. 

Notice that the overall central charge of the boundary VOA is $-2N +2(N-4) +2-4\times 2 = -14$, which coincides with the 
central charge of $\mathrm{su}(N|N-4)_{-2}\times \mathrm{su}(2)_1$. This suggests the boundary VOA will be an extension of this 
product of vertex algebras, which itself is a quotient of the product of $\mathfrak{su}(N|N-4)_{-2}\times \mathfrak{su}(2)_1$
Kac-Moody algebras. 

Operators involving an $\epsilon^{\alpha \beta}$ or $\epsilon_{\alpha \beta}$
tensors, of the schematic form $XX$, $X\chi$, $\chi \chi$, etc. need to be further dressed by $\tilde \psi^n$
in order to be gauge invariant. These operators transform as dimension $3/2$ anti-symmetric 2-index tensors of $\mathrm{su}(N|N-4)_{-2}$
which are doublets under the extra $SU(2)_1$ rotating the $n$ index of the extra fermions. They are also charged under $U(1)_C$.
Another operator of opposite $U(1)_C$ charge arises from bilinears of $Y$, $\psi$ dressed by $\tilde \chi$. 

It is reasonable to expect these extra currents will generate the extension of $\mathrm{su}(N|N-4)_{-2}\times \mathrm{su}(2)_1$
to the boundary VOA.

The case $N=4$ is special, as $U(1)_C$ should be enhanced to $SU(2)_C$ in the bulk. The operators of the form $\tilde \psi XX$
transform in antisymmetric fundamental tensors of $SU(4)_{-2}$. The operators of the form $\tilde \chi YY$
transform in antisymmetric anti-fundamental tensors of $SU(4)_{-2}$. These representations coincide, and could be rotated into each other 
by an enhanced $SU(2)_C$ outer automorphism. Thus the VOA appears to enjoy the same symmetry enhancement as the bulk QFT. 

All in all, $\mathfrak{A}_H[U(2),\C^{16},\C^4(2)]$ includes an $\mathfrak{su}(4)_{-2}\sim \mathfrak{so}(6)_{-2}$ current algebra, 
an $\mathrm{su}(2)_1$ current algebra and spin $3/2$ fields transforming as vectors of $\mathfrak{so}(6)_{-2}$, doublets of $\mathrm{su}(2)_1$
and doublets of $SU(2)_C$.

\subsection{A boundary VOA for $T[SU(3)]$}
The three-dimensional gauge theory which flows to $T[SU(3)]$
is a $U(1)\times U(2)$ gauge theory coupled to bifundamental hypermultiplets and $3$ fundamentals 
of $U(2)$ \cite{Gaiotto:2016aa}. 

The corresponding system of symplectic bosons includes the $U(1)$-charged $U(2)$ doublet 
$X^a$, $Y_a$ and the three extra $U(2)$ doublets $X_a^i$, $Y^b_j$. 

The level of the total $\mathfrak{su}(2)$ currents in $\mathfrak{u}(2)$ is $-4$, which is precisely what is needed for
anomaly cancellation. We only need to worry about the levels of the $\mathfrak{u}(1)$ current $J_1$ and the diagonal 
$\mathfrak{u}(1)$ $J_2$ current in $\mathfrak{u}(2)$:
\begin{align}
J_1 &= X^a Y_a \cr
J_2 &= \frac{1}{2}( X_a^i Y^a_i - X^a Y_a)
\end{align}
We have OPEs
\begin{align}
J_1(z) J_1(w) &\sim \frac{-2}{(z-w)^2} \cr
J_1(z) J_2(w) &\sim \frac{1}{(z-w)^2} \cr
J_2(z) J_2(w) &\sim \frac{-2}{(z-w)^2} 
\end{align}
Notice the resemblance to a Cartan matrix for $SU(3)$. 

In order to correct that anomaly with a well-defined set of boundary degrees of freedom, we include 
three complex fermions $\chi^1$, $\psi_1$,  $\chi^2$, $\psi_2$ and $\chi^3$, $\psi_3$. 
We will define the shifted total currents 
\begin{align}
J^t_1 &= X^a Y_a + \chi^1 \psi_1 - \chi^2 \psi_2 \cr
J^t_2 &= \frac{1}{2}( X_a^i Y^a_i - X^a Y_a) + \chi^2 \psi_2 - \chi^3 \psi_3
\end{align}
with no anomaly. The bulk Coulomb branch $\left(U(1)\times U(1)\right)_o$ symmetry is identified with the global part of the $U(1)$ symmetries 
acting on the complex fermions.

Thus our proposed boundary VOA is the $\mathfrak{u}(2)\times \mathfrak{u}(1)$-BRST quotient of the the VOA 
of eight symplectic bosons and three complex fermions. It has central charge $-15$. 

The $\mathfrak{su}(3)_{-2}$ currents 
\begin{equation}
J_{SU(3)} = X_a^i Y^a_j - \frac{\delta^i_j}{3} X_a^k Y^a_k
\end{equation} 
are obviously BRST closed. 

We also have an additional BRST closed $\mathfrak{u}(1)_3$ current together with two vertex operators 
built from the same current:
\begin{align}
\tilde J &= \chi^1 \psi_1 + \chi^2 \psi_2 +\chi^3 \psi_3 \cr
\tilde O^+ &= \chi^1 \chi^2 \chi^3 \cr
\tilde O^- &= \psi_1 \psi_2 \psi_3 
\end{align}
These are analogue to the $\mathrm{su}(2)_1$ generators in the $T[SU(2)]$ boundary VOA. 

The current algebra $\mathfrak{su}(3)_{-2} \times \mathfrak{u}(1)_3$ has central charge $-15$. It is reasonable to assume the 
full VOA is an extension of that current algebra. 

We can find three natural BRST-closed operators transforming in a fundamental of $\mathfrak{su}(3)_{-2}$ with charge $-1$ under 
$\mathrm{u}(1)_3$:
\begin{align}
O^i_1 &= \psi_1 X^a X_a^i \cr
O^i_2 &= \psi_2 \epsilon^{ab}Y_b X_a^i \cr
O^i_3 &= \psi_3 \epsilon^{ijk} \epsilon_{ab}Y^a_j Y^b_k 
\end{align}
These three operators all have dimension $\frac32$. They have $\left(U(1)\times U(1)\right)_o$ charges 
which precisely agree with a potential promotion of $\left(U(1)\times U(1)\right)_o$ to an $SU(3)_o$
which would make them into an $SU(3)_o$ triplet $O^i_A$. Another dual triplet $O_i^A$ can be built in the same manner using $\chi$ fermions. 

The OPE of $O^i_A$ and $O^j_B$ contain another set of anti-fundamental operators of dimension $2$ which
we can denote as $O^{[ij]}_{[AB]}$, distinct from $O_i^A$: 
\begin{align}
O^{[ij]}_{[12]} &= \psi_1 \psi_2 \epsilon^{ab} X_a^i X_b^j \cr
O^{[ij]}_{[23]} &= \psi_2 \psi_3\epsilon^{ijk} Y_b Y^b_k \cr
O^{[ij]}_{[31]} &= \psi_3 \psi_1\epsilon^{ijk} \epsilon_{ab}X^a Y^b_k 
\end{align}
A set of fundamental operators $O_{[ij]}^{[AB]}$ of dimension $2$ can be defined in a similar manner using $\chi$ fermions. 

The $O^i_A$ and $O_{[ij]}^{[AB]}$ operators are related by the action of $\tilde O^\pm$ and 
so are $O_i^A$ and $O^{[ij]}_{[AB]}$.

We can denote as $\mathrm{u}(1)_3$ 
the vertex algebra defined by $\mathfrak{u}(1)_3$ together with the associated vertex operators 
of charge $q$ and dimension $\frac32 q^2$. This has modules $M_i[\mathrm{u}(1)_3]$ formed by the vertex operators 
of charge $q + \frac{i}{3}$. 

We may conjecture that the operators above generate the full boundary VOA,
as a conformal extension of $\mathfrak{su}(3)_{-2} \times \mathrm{u}(1)_3$.

We observe that the character decomposes accordingly as 
\begin{equation}
\chi= \sum_{\lambda}^\infty \chi_{SU(3)_C}^{(\lambda)}\chi_{\mathfrak{su}(3)_{-2}}^{(\lambda)}\chi_{\mathrm{u}(1)_{3}}^{(\lambda \,\mathrm{mod} \,3)}
\end{equation}
where $\lambda$ are weights of $\mathfrak{su}(3)$ and $\lambda \,\mathrm{mod} \,3$ uses the identification of the weight modulo root lattice with the center $\Z_3$. 

\subsection{A VOA for $T[SU(N)]$}
The three-dimensional gauge theory which flows to $T[SU(N)]$
is a linear quiver, with $U(1)\times U(2)\times \cdots \times U(N-1)$ gauge fields coupled to bifundamental hypermultiplets and $N$ fundamentals 
of $U(N-1)$. 

We can denote the symplectic bosons between the $i$-th and $(i+1)$-th nodes 
as matrices ${\bf X}_i$ and ${\bf Y}_i$. We will denote as ${\bf \epsilon}_i$
the $\epsilon$ tensor at the $i$-th node and omit indices when contractions are 
unique. 

The level of the total $\mathfrak{su}(n)$ currents in $\mathfrak{u}(n)$ at each node is $-2n$, which is precisely what is needed for
anomaly cancellation. We only need to worry about the levels of the $\mathfrak{u}(1)$ currents $J_n$, diagonal 
components in $\mathfrak{u}(n)$:
\begin{equation}
J_n = \frac{1}{n}( {\bf X}_n \cdot {\bf Y}_n - {\bf Y}_{n-1} \cdot {\bf X}_{n-1})
\end{equation}
We have non-trivial OPEs
\begin{align}
J_n(z) J_n(w) &\sim \frac{-2}{(z-w)^2} \cr
J_n(z) J_{n+1}(w) &\sim \frac{1}{(z-w)^2} 
\end{align}
Notice the resemblance to a Cartan matrix for $SU(N)$. 

In order to correct that anomaly with a well-defined set of boundary degrees of freedom, we include 
$N$ complex fermions $\chi^i$, $\psi_i$, $i=1, \cdots N$. 
We will define the shifted total currents 
\begin{align}
J^t_n = \frac{1}{n}\mathrm{Tr}( {\bf X}_n \cdot {\bf Y}_n - {\bf Y}_{n-1} \cdot {\bf X}_{n-1})+ \chi_n \psi_n - \chi_{n+1} \psi_{n+1}
\end{align}
with no anomaly. 

The bulk Coulomb branch $U(1)^{N-1}_C$ symmetry is identified with the global part of the $U(1)$ symmetries 
acting on the complex fermions.

Thus we propose to take the $U(1)^{N-1}$-BRST quotient of the the above combination
of symplectic bosons and complex fermions.  

The $\mathfrak{su}(N)_{1-N}$ currents 
\begin{equation}
J_{SU(N)} =  {\bf X}_{N-1} \cdot {\bf Y}_{N-1} - \frac{1}{N} \mathrm{Tr}{\bf X}_{N-1} \cdot {\bf Y}_{N-1} 
\end{equation} 
are obviously BRST closed. 

We also have an additional BRST closed $\mathfrak{u}(1)_N$ current together with two vertex operators 
built from the same current:
\begin{align}
\tilde J = \chi^i \psi_i  \cr
\tilde O^+ = \prod_i \chi^i \cr
\tilde O^- = \prod_i \psi_i
\end{align}
These are analogue to the $\mathrm{su}(2)_1$ generators in the $T[SU(2)]$ case. 

We can find $N$ natural BRST-closed operators transforming in a fundamental of $\mathfrak{su}(N)_{1-N}$ with charge $-1$ under 
$\mathfrak{u}(1)_N$:
\begin{align}
{\bf O}_1 &= \psi_1 {\bf X}_1 \cdot {\bf X}_2 \cdots {\bf X}_{N-1}\cr
{\bf O}_2 &= \psi_2 ({\bf \epsilon}_2 \cdot {\bf Y}_1) \cdot {\bf X}_2 \cdots {\bf X}_{N-1}\cr
{\bf O}_3 &= \psi_3 (\epsilon_3 \cdot ({\bf Y}_2 \wedge {\bf Y}_2) \cdot \epsilon_2) \cdot {\bf X}_3 \cdots {\bf X}_{N-1} \cr
\cdots &= \cdots
\end{align}
These operators all have dimension $\frac{N}{2}$. They have $U(1)^{N-1}_C$ charges 
which precisely agree with a potential promotion of $U(1)^{N-1}_C$ to an $SU(N)_C$
which would make them into a $SU(N)_C\times \mathfrak{su}(N)_{1-N}$ bi-fundamental multiplet $O^i_A$
of charge $-1$ under $\mathfrak{u}(1)_N$. 

The OPE of multiple $O^i_A$ will contain operators $O^{[i_1 \cdots i_n]}_{[A_1 \cdots A_n]}$ involving $n$ of the $\psi_i$, 
transforming in antisymmetric powers of the fundamentals of $SU(N)_o\times \mathrm{su}(N)_{1-N}$.

Dual operators $O_{[i_1 \cdots i_n]}^{[A_1 \cdots A_n]}$ can be built in the same manner, but are obtained from 
the previous set by action of $\tilde O^\pm$. 

We may conjecture that the operators above generate the full boundary current algebra. 

We expect the character to decomposes as 
\begin{equation}
\chi= \sum_{\lambda}^\infty \chi_{SU(N)_C}^{(\lambda)}\chi_{\mathrm{su}(N)_{1-N}}^{(\lambda)}\chi_{\mathrm{u}(1)_{N}}^{(\lambda \,\mathrm{mod} \,N)}
\end{equation}
where $\lambda$ are weights of $\mathfrak{su}(N)$ and $\lambda \,\mathrm{mod} \,N$ uses the identification of the weight modulo root lattice with the center $\Z_N$. 

\section{The C-twist of standard $\CN=4$ gauge theories} \label{sec:C}
Consider a standard ${\cal N}=4$ gauge theory with gauge group $G$, matter fields in a symplectic representation $M$.

Upon C-twist, the bulk topological field theory can be identified with a Chern-Simons theory \cite{Companion}
based on a Lie algebra 
\begin{equation}
\mathfrak{l} = \mathfrak{g} \oplus \mathfrak{g}^* \oplus \Pi M
\end{equation}
with non-trivial brackets 
\begin{align}
[t_a,t_b] &= f_{ab}^c t_c \cr
[t_a,\tilde t^b] &= f_{ac}^b \tilde t^c \cr
[t_a,m^i] &= T_{a}^{ik} m^j \omega_{kj} \cr
\{m^i, m^j \} &= T^{ij}_a \tilde t^a
\end{align}
and level
\begin{equation}
K_{ab} = k_{ab} \qquad \qquad K^a_b = \delta^a_b \qquad \qquad K^{ij} = \omega^{ij}
\end{equation}
We included a possible one-loop shift $k_{ab}$ of the level for the compact part of the group. 
The level shift will be exactly opposite as the one encountered in the H-twist: $- 2 h$ from the 
gauge multiplet fermions and positive matter contributions to the boundary 't Hooft anomaly. 
\footnote{We can check that this level satisfies the appropriate constraints:
\begin{align}
f_{a[b}^d k_{c]d} &=0 \cr
f_{ab}^d \delta_d^c &= f^c_{ad} \delta^d_b \cr
T_{a}^{ik} \omega_{kt}\omega^{tj} &= T_{a}^{jk} \omega_{kt}\omega^{ti} = T^{ij}_d \delta^d_a \cr
\end{align}}

The gauge group is a bundle over the compact form of the gauge group. 

\subsection{Boundary conditions and VOA}
The simplest boundary condition we can conjecture being deformable consists of 
Dirichlet boundary conditions for both the gauge fields and the vectormultiplet scalars. This corresponds to 
a standard WZW boundary condition for the bulk Chern-Simons theory \cite{Companion}. 

As in more familiar situations, the boundary VOA $\mathfrak{A}_C[G,M,0]$ should be, essentially by definition, the WZW model 
associated to $\mathfrak{l}_K$.

A WZW model current algebra is not quite the same as the Kac-Moody algebra, even in the usual case of compact unitary gauge group: 
\begin{itemize}
\item The null vectors of the Kac-Moody algebra are removed.
\item Extra integrable modules for the Kac-Moody algebra are added in when the group is not simply connected. 
The modules are labelled by characters of the gauge group. 
\end{itemize}
Both the removal of null vectors and the extension by additional modules can
be interpreted as the contribution of boundary monopole operators to the 
boundary VOA. 

For example, a $U(1)_1$ Chern-Simons theory should support a chiral free fermion at a WZW boundary. 
This is an extension of a $\mathfrak{u}(1)_1$ current algebra by modules of integral charge. 
We can denote the extension as $\mathrm{u}(1)_1$.

Similarly, an $U(1)_2$ Chern-Simons theory should support an $\mathrm{u}(1)_2 \simeq \mathrm{su}(2)_1$ WZW model at a WZW boundary. 
This is an extension of a $\mathfrak{u}(1)_2$ current algebra by modules of even integral charge; and so on.  

We expect the same to happen for the WZW model associated to $\mathfrak{l}_K$.
Half-index calculations allow us to write down the character of such WZW models, but 
not to derive the precise form of the VOA. 

A more careful analysis presents the VOA as the Dolbeault homology of the {\it affine Grassmanian},
valued in certain bundles associated to the $m$ and $\tilde t$ generators of the Lie algebra  \cite{Companion}. 
It should be possible to fully compute the VOA structure from such definition. We leave it to future work. 

Dirichlet boundary conditions for the gauge theory can be modified to Nahm pole boundary conditions, 
where the gauge multiplet scalars diverge at the boundary as some reference solutions to Nahm equations. 
These should descend to ``oper-like'' boundary conditions for the CS theory. 

Again, we leave a discussion of the boundary VOA for these boundary conditions to future work.
Brane constructions suggest that these boundary conditions will play an important role in 
mirror symmetry.

\subsection{Half-index calculations}
If we ignored monopole contributions, the half-index for Dirichlet b.c. would simply be 
\begin{equation}
II^0_C(q; y) = \frac{\prod_{\alpha \in w(M)} \prod_{n>0}(1-y^\alpha q^{n})}{\prod_{\alpha \in w(\mathfrak{g})} \prod_{n>0} (1-y^\alpha q^{n})^2}
\end{equation}

The addition of monopole sectors modifies that to 
\begin{equation}
II^0_C(q; y;s) = \sum_{\mu \in \Lambda_w}  s^{\mu_{ab}} (-q^{\frac12})^{k(\mu,\mu)}\frac{\prod_{\alpha \in w(M)} \prod_{n>0}(1-y^\alpha q^{n+(\mu, \alpha)})}{\prod_{\alpha \in w(\mathfrak{g})} \prod_{n>0} (1-y^\alpha q^{n+(\mu, \alpha)})^2}
\end{equation}
where $k$ is the quadratic form which encodes the boundary 't Hooft anomaly for $G$ and $s$ a fugacity for the $U(1)_t$ charges
in case the gauge group has Abelian factors. One can recover this formula as a localization formula 
over the affine Grassmanian  \cite{Companion}. 

It is not hard to test some simple cases of this formula. For example, applied to a $U(1)$ gauge theory coupled to a single hypermultiplet 
of charge $1$, it gives an answer 
\begin{equation}
\sum_{m=-\infty}^\infty s^m (-q^{\frac12})^{m^2} \prod_{n>0} \frac{(1-q^{n+1-m} y)(1-q^{n+1+m} y)}{(1-q^{n+1})^2}
\end{equation}
which coincides with the vacuum character of a simple VOA: $\Sb \times \Ff$. This is reasonable: 
it indicates that Dirichlet b.c. are mirror to Neumann b.c. for the mirror hyper, dressed by a decoupled free complex fermion
so to produce the expected $\mathrm{u}(1|1)_1$ boundary currents. 

Recall that the mirror of an Abelian gauge theory with $n$ gauge fields and $N$ hypermultiplets and an $n \times N$ matrix of integral charges $Q$
is mirror to a gauge theory with matrix of charges $Q^\perp$. The statement holds if the gauge charges are minimal, 
i.e. if we can find an $(N-n) \times N$ matrix $q$ of integral ``flavor charges'' such that $\det (Q, q)=1$. Then the mirror charges are $(q^!, Q^!) = (Q, q)^{-1}$.

The simplest H-twisted boundary condition adds a Fermi multiplet for each hypermultiplet, leading to a $\mathfrak{u}(1)^n$ BRST quotient of $(\Sb \times \Ff)^N$.
The resulting VOA has $N$ pairs of fermionic currents produced as gauge-invariant bilinears of fermions and symplectic bosons of opposite charges. 
It also has $(N-n)$ pairs of bosonic currents. Inspection and index calculations strongly suggest an identification of the coset VOA with the C-twist
of Dirichlet b.c. for the mirror theory. 

\subsection{Relation to constructions in GL-twisted four-dimensional gauge theory}
It is possible to lift pure 3d ${\cal N}=4$ gauge theory to a configuration of four-dimensional 
${\cal N}=4$ gauge theory compactified on a segment, with Neumann boundary conditions. 

The four-dimensional theory has a continuous family of twists, parameterized by a ``topological gauge coupling''
$\Psi$ \cite{Kapustin:2006pk}. The C-twist of the 3d theory lifts to the $\Psi \to \infty$ limit of the four-dimensional twisted theory. 
The configuration with finite $\Psi$, though, make sense and can be considered a further deformation 
of the 3d TFT. 

Some boundary conditions for the 3d theory can also be lifted to the four-dimensional setup, 
by considering a half-strip configuration, with Neumann boundary conditions on the semi-infinite sides 
and Dirichlet of Nahm boundary conditions at the finite side. Appropriate junctions will have to be 
selected at the corners of the strip \cite{Gaiotto:2017euk,Creutzig:2017uxh}.

The configuration with Nahm boundary conditions is well understood. The setup will support a 
vertex algebra which is an extension of  a product $W^\mathfrak{g}_{\Psi-h} \times W^\mathfrak{g}_{-\Psi-h}$
of two W-algebras, obtained as a Drinfeld-Sokolov reduction of $\mathfrak{g}$ Kac-Moody algebras.

The extension consists of operators associated to finite segments of boundary monopole (aka 'tHooft) lines
at the Nahm boundary. They take the form of certain products of degenerate modules for 
$W^\mathfrak{g}_{\Psi-h} \times W^\mathfrak{g}_{-\Psi-h}$. These modules have arbitrarily negative dimension, 
which makes the boundary VOA unwieldy. 

If we employ Dirichlet boundary conditions, we have some extension/modification of a product of Kac-Moody algebras 
$\mathfrak{g}_{\Psi-h} \times \mathfrak{g}_{-\Psi-h}$. If we ignore the extension, it is easy to make contact 
with our boundary VOAs: the diagonal combination of these currents gives the level $-2 h$ currents for the $t$ generators, 
while the anti-diagonal goes in the $\Psi \to \infty$ limit to the $\tilde t$ generators. 

The structure of the extension is not well understood. It involves a category of modules labelled by 
D-modules on the affine Grassmanian $\mathrm{Gr}_G$ \cite{FG}. It should be possible to make contact with 
the three-dimensional construction involving homology on the affine Grassmanian.

The lift to four dimensions and deformation to finite $\Psi$ is possible in the presence of matter as well, 
but only if the matter can be organized into two representations $M^{(1)}$ and $M^{(2)}$ 
which can be used as fermionic generators to extend a $\mathfrak{g}$ to two super-algebras 
$\mathfrak{g}^{(1)}$ and $\mathfrak{g}^{(2)}$. 

Then the appropriate VOAs are built as above from Kac-Moody algebras (or associated W-algebras) of the form 
$\mathfrak{g}^{(1)}_{\Psi-h^{(1)}} \times \mathfrak{g}^{(2)}_{-\Psi-h^{(2)}}$. 
In the limit $\Psi \to \infty$, it is easy to make contact with theLie super-algebra we 
employed in the discussion of the C-twist. 

\section{Open questions and other speculations}
We may conclude with a few extra open problems:
\begin{enumerate}
\item There are a variety of ``exotic'' 3d $\CN=4$ theories \cite{Gaiotto:2008sd,Hosomichi:2008jd},
and Chern-Simons theories with even more supersymmetry, such as the ABJM theory \cite{Aharony:2008ug}.
These theories can be twisted \cite{Kapustin:2009cd} and may have interesting holomorphic boundary conditions. 
It would be nice to study them. 

\item Two-dimensional systems with $(0,4)$ supersymmetry are quite constrained. It is tricky, but possible \cite{Tong:2014yna}, to write down 
interactions involving the basic 2d $(0,4)$ supermultiplets, called $(0,4)$ Fermi multiplets, $(0,4)$ hypermultiplets
and $(0,4)$ twisted hypermultiplets. Similar couplings also exist when the 2d $(0,4)$ hypermultiplets and twisted hypermultiplets are replaced by boundary values of 3d $\CN=4$ multiplets. 

Of course, it is not obvious that such boundary conditions should be deformable. We can imagine, though, 
a speculative setup where $(0,4)$ 2d twisted hypermultiplets combine with the bulk vectormultiplets 
to give a VOA coset involving some super-Lie algebra $\hat{g}$ which extends $g$
by some fermionic generators originating from the 2d twisted hypermultiplets.

The bulk hypermultiplet representation $M$ should also be extended to a representation $\hat M$ of $\hat g$,
involving symplectic bosons in $M$ plus extra complex fermions fro the $(0,4)$ Fermi multiplets. 
The auxiliary 2d theory $A_{2d}$ should also include a $\hat g$ current algebra in order for the 
$\hat g$ BRST reduction to make sense. 

\end{enumerate}

\noindent {\bf Acknowledgements.} K.C. and D.G. are supported by the NSERC
Discovery Grant program and by the Perimeter Institute for Theoretical
Physics. Research at Perimeter Institute is supported by the
Government of Canada through Industry Canada and by the Province of
Ontario through the Ministry of Research and Innovation.


\begin{thebibliography}{10}




\bibitem{Vafa:1994tf}
C.~Vafa and E.~Witten, \emph{{A Strong coupling test of S duality}},
  \href{http://dx.doi.org/10.1016/0550-3213(94)90097-3}{\emph{Nucl. Phys.} {\bf
  B431} (1994) 3--77}, [\href{http://arxiv.org/abs/hep-th/9408074}{{\tt
  hep-th/9408074}}].

\bibitem{Nekrasov:2003rj}
N.~Nekrasov and A.~Okounkov, \emph{{Seiberg-Witten theory and random
  partitions}}, \href{http://dx.doi.org/10.1007/0-8176-4467-9_15}{\emph{Prog.
  Math.} {\bf 244} (2006) 525--596},
  [\href{http://arxiv.org/abs/hep-th/0306238}{{\tt hep-th/0306238}}].

\bibitem{Alday:2009aq}
L.~F. Alday, D.~Gaiotto and Y.~Tachikawa, \emph{{Liouville Correlation
  Functions from Four-dimensional Gauge Theories}},
  \href{http://dx.doi.org/10.1007/s11005-010-0369-5}{\emph{Lett. Math. Phys.}
  {\bf 91} (2010) 167--197}, [\href{http://arxiv.org/abs/0906.3219}{{\tt
  0906.3219}}].

\bibitem{Beem:2014kka}
C.~Beem, L.~Rastelli and B.~C. van Rees, \emph{W symmetry in six dimensions},
  \href{http://dx.doi.org/10.1007/JHEP05(2015)017}{\emph{JHEP} {\bf 05} (2015)
  017}, [\href{http://arxiv.org/abs/1404.1079}{{\tt 1404.1079}}].

\bibitem{Beem:2013aa}
C.~Beem, M.~Lemos, P.~Liendo, W.~Peelaers, L.~Rastelli and B.~C. van Rees,
  \emph{Infinite chiral symmetry in four dimensions},
  \href{http://arxiv.org/abs/1312.5344}{{\tt 1312.5344}}.

\bibitem{Beem:2014aa}
C.~Beem, W.~Peelaers, L.~Rastelli and B.~C. van Rees, \emph{Chiral algebras of
  class s},  \href{http://arxiv.org/abs/1408.6522}{{\tt 1408.6522}}.

\bibitem{Gaiotto:2017euk}
D.~Gaiotto and M.~Rap{\v c}{\'a}k, \emph{{Vertex Algebras at the Corner}},
  \href{http://arxiv.org/abs/1703.00982}{{\tt 1703.00982}}.

\bibitem{Gaiotto:2016wcv}
D.~Gaiotto, \emph{{Twisted compactifications of 3d N = 4 theories and conformal
  blocks}},  \href{http://arxiv.org/abs/1611.01528}{{\tt 1611.01528}}.

\bibitem{Kapustin:2006pk}
A.~Kapustin and E.~Witten, \emph{{Electric-Magnetic Duality And The Geometric
  Langlands Program}},
  \href{http://dx.doi.org/10.4310/CNTP.2007.v1.n1.a1}{\emph{Commun. Num. Theor.
  Phys.} {\bf 1} (2007) 1--236},
  [\href{http://arxiv.org/abs/hep-th/0604151}{{\tt hep-th/0604151}}].

\bibitem{Gaiotto:2008ak}
D.~Gaiotto and E.~Witten, \emph{{S-Duality of Boundary Conditions In N=4 Super
  Yang-Mills Theory}},
  \href{http://dx.doi.org/10.4310/ATMP.2009.v13.n3.a5}{\emph{Adv. Theor. Math.
  Phys.} {\bf 13} (2009) 721--896}, [\href{http://arxiv.org/abs/0807.3720}{{\tt
  0807.3720}}].

\bibitem{Gaiotto:2016aa}
D.~Gaiotto, \emph{S-duality of boundary conditions and the geometric langlands
  program},  \href{http://arxiv.org/abs/1609.09030}{{\tt 1609.09030}}.

\bibitem{Creutzig:2017uxh}
T.~Creutzig and D.~Gaiotto, \emph{{Vertex Algebras for S-duality}},
  \href{http://arxiv.org/abs/1708.00875}{{\tt 1708.00875}}.

\bibitem{Witten:1988ze}
E.~Witten, \emph{{Topological Quantum Field Theory}},
  \href{http://dx.doi.org/10.1007/BF01223371}{\emph{Commun. Math. Phys.} {\bf
  117} (1988) 353}.

\bibitem{Witten:1988xj}
E.~Witten, \emph{{Topological Sigma Models}},
  \href{http://dx.doi.org/10.1007/BF01466725}{\emph{Commun. Math. Phys.} {\bf
  118} (1988) 411}.

\bibitem{Rozansky:1996bq}
L.~Rozansky and E.~Witten, \emph{{HyperKahler geometry and invariants of three
  manifolds}}, \href{http://dx.doi.org/10.1007/s000290050016}{\emph{Selecta
  Math.} {\bf 3} (1997) 401--458},
  [\href{http://arxiv.org/abs/hep-th/9612216}{{\tt hep-th/9612216}}].

\bibitem{Kapustin:2010ag}
A.~Kapustin and K.~Vyas, \emph{{A-Models in Three and Four Dimensions}},
  \href{http://arxiv.org/abs/1002.4241}{{\tt 1002.4241}}.

\bibitem{Kapustin:2008sc}
A.~Kapustin, L.~Rozansky and N.~Saulina, \emph{{Three-dimensional topological
  field theory and symplectic algebraic geometry I}},
  \href{http://dx.doi.org/10.1016/j.nuclphysb.2009.01.027}{\emph{Nucl. Phys.}
  {\bf B816} (2009) 295--355}, [\href{http://arxiv.org/abs/0810.5415}{{\tt
  0810.5415}}].

\bibitem{Kapustin:2009uw}
A.~Kapustin and L.~Rozansky, \emph{{Three-dimensional topological field theory
  and symplectic algebraic geometry II}},
  \href{http://dx.doi.org/10.4310/CNTP.2010.v4.n3.a1}{\emph{Commun. Num. Theor.
  Phys.} {\bf 4} (2010) 463--550}, [\href{http://arxiv.org/abs/0909.3643}{{\tt
  0909.3643}}].

\bibitem{Bullimore:2016nji}
M.~Bullimore, T.~Dimofte, D.~Gaiotto and J.~Hilburn, \emph{{Boundaries, Mirror
  Symmetry, and Symplectic Duality in 3d $\mathcal{N}=4$ Gauge Theory}},
  \href{http://dx.doi.org/10.1007/JHEP10(2016)108}{\emph{JHEP} {\bf 10} (2016)
  108}, [\href{http://arxiv.org/abs/1603.08382}{{\tt 1603.08382}}].

\bibitem{Chung:2016pgt}
H.-J. Chung and T.~Okazaki, \emph{{(2,2) and (0,4) supersymmetric boundary
  conditions in 3d $\mathcal{N}$ = 4 theories and type IIB branes}},
  \href{http://dx.doi.org/10.1103/PhysRevD.96.086005}{\emph{Phys. Rev.} {\bf
  D96} (2017) 086005}, [\href{http://arxiv.org/abs/1608.05363}{{\tt
  1608.05363}}].

\bibitem{Dimofte:2017tpi}
T.~Dimofte, D.~Gaiotto and N.~M. Paquette, \emph{{Dual Boundary Conditions in
  3d SCFT's}},  \href{http://arxiv.org/abs/1712.07654}{{\tt 1712.07654}}.

\bibitem{Companion}
K.~Costello, T.~Dimofte and D.~Gaiotto, \emph{Boundary vertex algebras and
  holomorphic twists}, {\emph{To Appear} }.


\bibitem{Witten:1988hf}
E.~Witten, \emph{{Quantum Field Theory and the Jones Polynomial}},
  \href{http://dx.doi.org/10.1007/BF01217730}{\emph{Commun. Math. Phys.} {\bf
  121} (1989) 351--399}.

\bibitem{Witten:2010cx}
E.~Witten, \emph{{Analytic Continuation Of Chern-Simons Theory}}, {\emph{AMS/IP
  Stud. Adv. Math.} {\bf 50} (2011) 347--446},
  [\href{http://arxiv.org/abs/1001.2933}{{\tt 1001.2933}}].

\bibitem{2016arXiv161100311B}
D.~{Butson} and P.~{Yoo}, \emph{{Degenerate Classical Field Theories and
  Boundary Theories}}, {\emph{ArXiv e-prints} (Nov., 2016) },
  [\href{http://arxiv.org/abs/1611.00311}{{\tt 1611.00311}}].

\bibitem{Imamura:2011su}
Y.~Imamura and S.~Yokoyama, \emph{{Index for three dimensional superconformal
  field theories with general R-charge assignments}},
  \href{http://dx.doi.org/10.1007/JHEP04(2011)007}{\emph{JHEP} {\bf 04} (2011)
  007}, [\href{http://arxiv.org/abs/1101.0557}{{\tt 1101.0557}}].

\bibitem{Kapustin:2011jm}
A.~Kapustin and B.~Willett, \emph{{Generalized Superconformal Index for Three
  Dimensional Field Theories}},  \href{http://arxiv.org/abs/1106.2484}{{\tt
  1106.2484}}.

\bibitem{Gadde:2013wq}
A.~Gadde, S.~Gukov and P.~Putrov, \emph{{Walls, Lines, and Spectral Dualities
  in 3d Gauge Theories}},
  \href{http://dx.doi.org/10.1007/JHEP05(2014)047}{\emph{JHEP} {\bf 05} (2014)
  047}, [\href{http://arxiv.org/abs/1302.0015}{{\tt 1302.0015}}].

\bibitem{Gadde:2013sca}
A.~Gadde, S.~Gukov and P.~Putrov, \emph{{Fivebranes and 4-manifolds}},
  \href{http://arxiv.org/abs/1306.4320}{{\tt 1306.4320}}.

\bibitem{VOAExt}
K.~Costello, T.~Creutzig and D.~Gaiotto, \emph{Higgs and coulomb branches from
  vertex operator algebras}, {\emph{To Appear} }.

\bibitem{Assel:2015oxa}
B.~Assel and J.~Gomis, \emph{{Mirror Symmetry And Loop Operators}},
  \href{http://dx.doi.org/10.1007/JHEP11(2015)055}{\emph{JHEP} {\bf 11} (2015)
  055}, [\href{http://arxiv.org/abs/1506.01718}{{\tt 1506.01718}}].

\bibitem{1990NuPhB.329..649K}
D.~{Karabali} and H.~J. {Schnitzer}, \emph{{BRST quantization of the gauged WZW
  action and coset conformal field theories}},
  \href{http://dx.doi.org/10.1016/0550-3213(90)90075-O}{\emph{Nuclear Physics
  B} {\bf 329} (Jan., 1990) 649--666}.

\bibitem{Nakajima:2015txa}
H.~Nakajima, \emph{{Towards a mathematical definition of Coulomb branches of
  $3$-dimensional $\mathcal{N}=4$ gauge theories, I}},
  \href{http://dx.doi.org/10.4310/ATMP.2016.v20.n3.a4}{\emph{Adv. Theor. Math.
  Phys.} {\bf 20} (2016) 595--669},
  [\href{http://arxiv.org/abs/1503.03676}{{\tt 1503.03676}}].

\bibitem{Braverman:2016wma}
A.~Braverman, M.~Finkelberg and H.~Nakajima, \emph{{Towards a mathematical
  definition of Coulomb branches of $3$-dimensional $\mathcal N=4$ gauge
  theories, II}},  \href{http://arxiv.org/abs/1601.03586}{{\tt 1601.03586}}.



\bibitem{Costello:2004ei} 
  K.~Costello,
  \emph{Topological conformal field theories and Calabi-Yau categories},
  Adv.\ Math.\  {\bf 210}, 165 (2007)
  doi:10.1016/j.aim.2006.06.004
  [math/0412149 [math-qa]].

\bibitem{Kapustin:2004df} 
  A.~Kapustin and L.~Rozansky,
  \emph{On the relation between open and closed topological strings},
  Commun.\ Math.\ Phys.\  {\bf 252}, 393 (2004)
  doi:10.1007/s00220-004-1227-z
  [hep-th/0405232].


\bibitem{Kapustin:1999ha}
A.~Kapustin and M.~J. Strassler, \emph{{On mirror symmetry in three-dimensional
  Abelian gauge theories}},
  \href{http://dx.doi.org/10.1088/1126-6708/1999/04/021}{\emph{JHEP} {\bf 04}
  (1999) 021}, [\href{http://arxiv.org/abs/hep-th/9902033}{{\tt
  hep-th/9902033}}].

\bibitem{FG}
E.~Frenkel and D.~Gaiotto, \emph{Quantum Langlands dualities of
  boundary conditions, D-modules, and conformal blocks}, {\emph{To Appear} }.

\bibitem{Gaiotto:2008sd}
D.~Gaiotto and E.~Witten, \emph{{Janus Configurations, Chern-Simons Couplings,
  And The theta-Angle in N=4 Super Yang-Mills Theory}},
  \href{http://dx.doi.org/10.1007/JHEP06(2010)097}{\emph{JHEP} {\bf 06} (2010)
  097}, [\href{http://arxiv.org/abs/0804.2907}{{\tt 0804.2907}}].

\bibitem{Hosomichi:2008jd}
K.~Hosomichi, K.-M. Lee, S.~Lee, S.~Lee and J.~Park, \emph{{N=4 Superconformal
  Chern-Simons Theories with Hyper and Twisted Hyper Multiplets}},
  \href{http://dx.doi.org/10.1088/1126-6708/2008/07/091}{\emph{JHEP} {\bf 07}
  (2008) 091}, [\href{http://arxiv.org/abs/0805.3662}{{\tt 0805.3662}}].

\bibitem{Aharony:2008ug}
O.~Aharony, O.~Bergman, D.~L. Jafferis and J.~Maldacena, \emph{{N=6
  superconformal Chern-Simons-matter theories, M2-branes and their gravity
  duals}}, \href{http://dx.doi.org/10.1088/1126-6708/2008/10/091}{\emph{JHEP}
  {\bf 10} (2008) 091}, [\href{http://arxiv.org/abs/0806.1218}{{\tt
  0806.1218}}].

\bibitem{Kapustin:2009cd}
A.~Kapustin and N.~Saulina, \emph{{Chern-Simons-Rozansky-Witten topological
  field theory}},
  \href{http://dx.doi.org/10.1016/j.nuclphysb.2009.07.006}{\emph{Nucl. Phys.}
  {\bf B823} (2009) 403--427}, [\href{http://arxiv.org/abs/0904.1447}{{\tt
  0904.1447}}].

\bibitem{Tong:2014yna}
D.~Tong, \emph{{The holographic dual of $AdS_{3} \times S^{3} \times S^{3}
  \times S^{1}$}}, \href{http://dx.doi.org/10.1007/JHEP04(2014)193}{\emph{JHEP}
  {\bf 04} (2014) 193}, [\href{http://arxiv.org/abs/1402.5135}{{\tt
  1402.5135}}].

\end{thebibliography}

\providecommand{\href}[2]{#2}\begingroup\raggedright\endgroup

\end{document}